\documentclass[12pt]{article}

\usepackage{amssymb}
\usepackage{amsmath}
\usepackage{graphicx}

\newcommand{\bq}{\begin{eqnarray}}
\newcommand{\eq}{\end{eqnarray}}
\newcommand{\bqn}{\begin{eqnarray*}}
\newcommand{\eqn}{\end{eqnarray*}}
\newcommand{\rr}{{\mathbf r}}
\newcommand{\QQ}{{\mathbf Q}}

\makeatletter
\@addtoreset{equation}{section} 
\makeatother

\renewcommand{\theequation}{\arabic{section}.\arabic{equation}}

\begin{document}
\begin{titlepage}

\title{\bf Stability boundaries, percolation threshold, and two phase
coexistence for polydisperse fluids of adhesive colloidal particles}

\author{{\bf Riccardo Fantoni}\footnote{e-mail: {\rm rfantoni@unive.it}},
{\bf Domenico Gazzillo}\footnote{e-mail: {\rm gazzillo@unive.it}}, and
{\bf Achille Giacometti}\footnote{e-mail: {\rm achille@unive.it}}\\
{\sl Istituto Nazionale per la Fisica della Materia and}\\
{\sl Dipartimento di Chimica Fisica, Universit\`a di Venezia,}\\ 
{\sl S. Marta DD 2137, I-30123 Venezia, Italy}}

\maketitle

\begin{abstract}
\noindent
We study the polydisperse Baxter model of sticky hard
spheres (SHS) in the modified Mean Spherical Approximation (mMSA). 
This closure is known to be the zero-order approximation (C0)
of the Percus-Yevick (PY) closure in a density expansion.
The simplicity of the closure allows a full analytical study
of the model. In particular we
study stability boundaries,
the percolation threshold, and the gas-liquid coexistence curves.
Various possible sub-cases of the model are treated in details.
Although the detailed behavior depends upon the particularly
chosen case, we find that, in general, polydispersity inhibits
instabilities, increases the extent of the non percolating
phase, and diminishes the size of the gas-liquid coexistence region. 
We also consider the first-order improvement of the mMSA (C0) closure
(C1) and compare the percolation and gas-liquid boundaries for the
one-component system with recent Monte Carlo simulations. 
Our results provide a qualitative understanding of the effect of
polydispersity on SHS models and are expected to shed new light on the
applicability of SHS models for colloidal mixtures.
\end{abstract}

\end{titlepage}

\section{Introduction}
\label{sec:intro}
In sterically stabilized colloidal mixtures, particles are coated
with polymer brushes to prevent irreversible flocculation due to
van der Waals attraction \cite{Lowen94}. If the solvent is a moderate one, 
a lowering of the temperature yields very strong attraction with a range 
much less than the typical colloidal size. In microemulsions of
polydispersed spherical water droplets each coated by a monolayer of
sodium di-2-ethylhexylsulfosuccinate dispersed in a continuum of oil,
the droplets interact with each other via a hard core plus a short
range attractive potential, the strength of which increases with
temperature \cite{Chen94}.
For these systems, a very useful theoretical model is the sticky
hard sphere (SHS) model proposed by Baxter \cite{Baxter68} long time
ago for atomic liquids. In the original Baxter solution
\cite{Baxter68,Watts71} the one-component Orstein-Zernike (OZ)
integral equation was analytically solved within the Percus-Yevick
(PY) approximation. Successive extension to mixtures \cite{Barboy79},
however, proved to be a formidable task in view of the fact that a
large (infinite
\footnote{Strictly speaking we should distinguish between {\sl
discrete polydispersity} (multicomponent mixture with a large number
of components $p\approx 10^2 \div 10^3$) and a {\sl continuous
polydispersity} corresponding to $p\to\infty$ with a continuous
distribution of sizes or other properties. This distinction will be
specified in more details in Section \ref{sec:binodal}.}) number of
coupled quadratic equations ought to be 
solved numerically in order to have a complete understanding of both
thermodynamics and structure of the model. This is the reason why, to
the best of our knowledge, only binary mixtures have been explicitly
discussed so far in this framework \cite{Barboy79}.
Moreover it has been proven by Stell \cite{Stell91} that sticky
spheres of equal diameter in the Baxter limit are not
thermodynamically stable and size polydispersity can be expected to
restore thermodynamic stability.

Motivated by this scenario, it was recently proposed \cite{Gazzillo04}
a simpler approximation (mMSA closure) having the advantage that also
the multicomponent case could be worked out analytically
\cite{Gazzillo00,Gazzillo03}. Further analysis and comparison with
both Monte Carlo (MC) and PY results
\cite{Gazzillo04,Miller03,Miller04} in the one-component case,  
have shown that the mMSA closure for Baxter model is a reliable one 
up to experimentally significant densities. The price to pay for this
simplification is that only the energy equation of state gives rise to
a critical behavior, the other two routes yielding either a non-critical
behavior (compressibility), or a diverging equation of state (virial).

In this work we pursue this investigation by studying the
multicomponent version of the model proposed in
Ref. \cite{Gazzillo04}, and analyzing various consequences. 
We first solve the multicomponent version of Baxter model within
the mMSA closure, and show that the solution is equivalent to the one
derived in Ref. \cite{Gazzillo00} for a companion SHS model. The
solution, derived in terms of an auxiliary function called Baxter
factor correlation, turns out to be formally similar to that derived
with the PY closure. However, and this is the crux of the matter, the
matrix function representing the stickiness parameters is
unconstrained, unlike the PY counterpart. In order to make further
progress and derive the multicomponent energy equation of state, a
further assumption is necessary on the matrix representing the
stickiness parameters. As discussed previously (see
Ref. \cite{Gazzillo00} for details) a remarkable simplification occurs
when the general element of this matrix has the form of a sum of dyads
(i.e. it is 
dyadic). In these cases the necessary matrix inversion can be carried
out analytically and all measurable quantities can then be
computed. Physically, this reduction to a dyadic form amounts to
assume a relation among polydispersity in size and polydispersity in 
stickiness, that is on the adhesion forces. In addition to the two
cases proposed in Ref. \cite{Gazzillo00} (denoted as Case I and II in the
following) and that proposed in Ref. \cite{Tutschka98} (Case IV), we
shall consider two further cases. The first one (Case III) is a
physically motivated variant of Case I, whereas the second one (Case
V) has its main justification in the simplifying features occurring
when one attempts to go beyond the mMSA closure with a density
perturbative approach (to first order this will be called C1, as in
Ref. \cite{Gazzillo04}, for reasons which will become apparent in the
rest of the paper). 

The main results of our analysis are the following. We derive the
instability curves in three of the considered cases (Case I-III)
within the mMSA approximation and analyze the effect of polydispersity
in some detail. In order to test the reliability of the mMSA
approximation, we also consider the first-order correction (C1) in the
one-component case and compare with the PY result. 

Next we consider the effect of polydispersity on the percolation
threshold. This is an interesting phenomenon on its own right and has
attracted considerable attention recently
\cite{Coniglio77,Chiew83,Chiew89,Miller03,Miller04}, being a
paradigmatic example of flocculation instability. In particular,
recent Monte Carlo simulations \cite{Miller03,Miller04} on
monodisperse (one-component) spheres with sticky adhesion have clearly
tested the performance of analytical calculations based on the PY
approximation \cite{Chiew83,Chiew89}. We then study the percolation
transition as a function of polydispersity in all above mentioned
cases within mMSA. Again we can discriminate the effect of
polydispersity on the percolation line, and also compare it with the
first-order correction C1, the PY approximation and MC simulations in
the one-component case. 

Next we consider phase equilibrium. A major obstacle to the analysis
of phase transition in polydisperse systems is posed by the fact that,
in principle, one has to deal with a large (infinite) number of
integral non-linear equations corresponding to the coexistence
conditions among various phases. In this model, however, as it also
occurs in other simpler models such as hard spheres (HS)
\cite{Sollich02}, van der Waals fluids \cite{Bellier00} and in more
complex cases such as factorizable hard-sphere Yukawa potentials
\cite{Kalyuzhnyi03,Kalyuzhnyi04}, the task can be  carried out in full
detail in view of the fact that the (excess) free energy depends upon
only a finite number of moments of the size distribution function. In
the particular case of two-phase coexistence, we derive the cloud and
shadow curves of all Cases in the mMSA approximation. We compare the
results with those derived earlier for a polydisperse van der Waals
fluid \cite{Bellier00}, and discuss analogies 
and differences in this respect. Finally we compare the results of the
mMSA one-component case with the first-order correction, the PY
approximation, and the results of MC simulations. 

The plan of the paper is as follows. In Section \ref{sec:baxter} we
define the multicomponent SHS model, give the solution for Baxter
factor correlation function in the mMSA (C0) approximation, and define
the various Cases of polydispersion models taken under exam; In
Section \ref{sec:c1} we give the solution for Baxter factor
correlation function in the C1 approximation and show how Case V is
particularly suitable to study the polydisperse system analytically;
in Section \ref{sec:inst} we analytically derive the instability
boundaries; in Section \ref{sec:perc} we find analytically the
percolation thresholds; In Section \ref{sec:binodal} we derive
numerically the two phase coexistence curves;
In Section \ref{sec:concl} we lay down our conclusions and further
developments.    
\section{Baxter model and modified MSA solution}
\label{sec:baxter}

In Baxter model of sticky hard spheres (SHS1), one starts adding to
the hard sphere (HS) potential a square-well tail with \cite{Perram75}
\begin{equation}
\phi _{ij}(r)=-k_{B}T\ln \left( \ \frac{1}{12\tau _{ij}}
\frac{R_{ij}}{R_{ij}-\sigma _{ij}}\right)~,\qquad \sigma _{ij}\leq
r\leq R_{ij}~, 
\label{m1}
\end{equation}
where $\sigma _{ij}=(\sigma _{i}+\sigma _{j})/2$ ($\sigma _{i}$ being the HS
diameter of species $i$), $R_{ij}-\sigma _{ij}$ denotes the well width, $%
k_{B}$ is Boltzmann constant, $T$ the temperature, and the dimensionless
parameter $\tau _{ij}^{-1}\geq 0$ measures the strength of surface
adhesiveness or `stickiness' between particles of species $i$ and $j$ ($\tau
_{ij}$ is also an unspecified increasing function of $T$). The sticky limit
corresponds to taking $\left\{ R_{ij}\right\} \rightarrow \left\{ \sigma
_{ij}\right\} $.

The Baxter form of the Ornstein-Zernike (OZ) integral equations for this
model admits a very simple analytic solution if one uses the following
modified Mean Spherical Approximation (mMSA)
\begin{equation}
c_{ij}\left( r\right) =f_{ij}(r)\text{ \ \ \ \ \ \ for\ \ \ }r\geq
\sigma_{ij}~,  \label{m2} 
\end{equation}
where $c_{ij}\left( r\right) $ and $f_{ij}(r)=\exp \left[ -\beta \phi
_{ij}(r)\right] -1$ are the direct correlation function and the Mayer
function, respectively [$\beta =(k_{B}T)^{-1}$].
This can be easily inferred by using the formalism introduced
in Ref. \cite{Gazzillo04}. As pointed out in that reference, the mMSA closure
can be reckoned as a zero-order approximation in a perturbative
expansion, and hence it will also be denoted as C0 henceforth.
In terms of Baxter factor correlation functions
$q_{ij}(r)$, its extension to mixtures reads 
\begin{equation}
q_{ij}(r)=\left\{ 
\begin{array}{ll}
\frac{1}{2}a_{i}(r-\sigma _{ij})^{2}+(b_{i}+a_{i}\sigma _{ij})(r-\sigma
_{ij})+K_{ij}~, & L_{ij}=(\sigma_i-\sigma_j)/2\leq r\leq \sigma _{ij}~, \\ 
0~,             & \text{elsewhere}~,
\end{array}
\right.  \label{m3}
\end{equation}

\begin{equation}
a_{i}=\frac{1}{\Delta }+\frac{3\xi _{2}\sigma _{i}}{\Delta ^{2}}-
\frac{12\zeta _{i}}{\Delta }~,\qquad b_{i}=\left( \frac{1}{\Delta
}-a_{i}\right)\frac{\sigma _{i}}{2}~,  \label{m4}
\end{equation}

\begin{equation}
\xi _{n}=\frac{\pi }{6}\sum_{i=1}^{p}\rho _{i}\sigma _{i}^{n}~,\qquad 
\zeta_{i}=\frac{\pi }{6}\sum_{m=1}^{p}\rho _{m}\sigma
_{m}K_{im}~,\qquad 
\Delta=1-\xi _{3}~,  \label{m5}
\end{equation}
with $p$ being the number of components, $\rho_i$ the number density
of species $i$, and 
\begin{equation}
K_{ij}^{(mMSA)}=\frac{1}{12\tau _{ij}}\sigma _{ij}^{2}\equiv
K^0_{ij}~.  \label{m6} 
\end{equation}

We remark that although Eqs. ($\ref{m3}$)-($\ref{m5}$) are formally
identical to their PY counterpart, this result is in fact simpler in such 
they differ in the quantity $K_{ij}$ which in the PY approximation reads
\cite{Perram75}
\begin{equation}
K_{ij}^{(PY)}=K^0_{ij}\,y_{ij}^{(PY)}(\sigma_{ij}) \equiv
\frac{1}{12}\lambda _{ij}\sigma _{ij}^{2}~, 
\label{m7}
\end{equation}
where $y_{ij}^{(PY)}(\sigma _{ij})$ is the contact value of the PY
cavity function. In general, the parameters $\lambda _{ij}$ can be
determined only numerically by solving a set of \ $p(p+1)/2$ coupled
quadratic equations \cite{Perram75,Barboy79}, and this makes the
multicomponent PY solution of limited interest from the practical viewpoint.
In particular a global analysis of the phase diagram proves to be a formidable
task within the PY approximation \cite{Barboy79}. On the other hand, in
view of the simplicity of Eq. (\ref{m6}) with respect to its PY counterpart
Eq. (\ref{m7}), this is indeed possible within the
mMSA (C0) approximation. The above results is, moreover, fully
equivalent to a parallel but different sticky HS model (SHS3) studied by
us in previous work 
\cite{Gazzillo00,Gazzillo03}. Hence, as discussed in those references, 
this analysis can be pursued analytically provided that $K_{ij}$ has a 
dyadic form. To this aim, we consider polydisperse fluids with
HS diameters distributed according to a Schulz distribution
\footnote{Here, for simplicity, we disregard possible complicancies
arising from the fact that unphysically large particles are included
in this analysis. These were discussed in Ref. \cite{Kalyuzhnyi03}.}. 

As regards stickiness, we choose to keep it either constant or related
to the particle size. There are two main reasons for this. First,
one expects the adhesion forces to depend upon the area of the contact
surface between two particles (see Fig. \ref{fig:p-cases}), 
and hence on their sizes. Second and more practical reason,
is that this is a simple way of obtaining the required factorization.
As the stickiness-size relation is not clearly understood, we consider
five different possibilities, denoted as Case I-V henceforth. 
The three simplest choices are
\bq \label{m8}
\frac{1}{\tau _{ij}}=\frac{1}{\tau }\ \frac{\left\langle \sigma
\right\rangle ^{2}}{\sigma _{ij}^{2}},\qquad &\Longrightarrow& \qquad \left[
K_{ij}^{(mMSA)}\right]_{{\rm Case\ I}}=\frac{1}{12\tau }\left\langle
\sigma \right\rangle^{2}~,\\ \label{m9}  
\frac{1}{\tau _{ij}}=\frac{1}{\tau }\ \frac{\sigma _{i}\sigma _{j}}{\sigma
_{ij}^{2}},\qquad &\Longrightarrow& \qquad \left[ K_{ij}^{(mMSA)}\right]_{
{\rm Case\ II}}\text{ }=\frac{1}{12\tau }\ \sigma _{i}\sigma _{j}~,\\
\label{modelIII}
\frac{1}{\tau _{ij}}=\frac{1}{\tau }\ \frac{\left\langle \sigma^2
\right\rangle}{\sigma _{ij}^{2}},\qquad &\Longrightarrow& \qquad \left[
K_{ij}^{(mMSA)}\right]_{{\rm Case\ III}}=\frac{1}{12\tau }\left\langle
\sigma^2 \right\rangle~.
\eq
where $\left\langle \sigma \right\rangle $ is the average HS diameter 
($\left\langle F\right\rangle \equiv \sum_{i}x_{i}F_{i}$, here
$x_i=\rho_i/\rho$ is the molar fraction of species $i$ with
$\rho=\sum_i\rho_i$ the total number density) and $\tau $ is
assumed to depend only on the temperature, while the remaining factor
in $\tau _{ij}^{-1}$ is a measure of stickiness strength and is
related to the particle sizes. The physical interpretation of these choices
is the following. In Case I the stickiness is assumed to be proportional
to the surface contact area of two colloidal particles having average
size $\left\langle\sigma\right\rangle$, whereas in Case II the
adhesion of each particle is linearly related to its size. Case III,
finally, is a variant of Case I where one considers an average
stickiness rather than the stickiness of an average particle.

In all these cases the $K_{ij}^{(mMSA)}$ matrix can be factorized  as 
\begin{equation}
K_{ij}^{(mMSA)}=Y_{i}Y_{j}~,  \label{m10}
\end{equation}
with $Y_{i}$ having dimensions of length ($Y_{i}=\left( \sqrt{12\tau }%
\right) ^{-1}\left\langle \sigma \right\rangle $, $Y_{i}=\left( \sqrt{%
12\tau }\right) ^{-1}\ \sigma _{i}$, and $Y_{i}=\left( \sqrt{12\tau }%
\right)^{-1}\left\langle \sigma^2 \right\rangle^{1/2}$ in Case I, II,
and III, respectively). Note that Case I and II have already been
exploited by us in previous work \cite{Gazzillo00}.

We also consider a case similar to that proposed by Tutschka and Kahl
\cite{Tutschka98} (henceforth denoted as Case IV)
\bq \label{modelIV}
\frac{1}{\tau_{ij}}=\frac{1}{\tau}~,
\eq
In this case the $K_{ij}^{(mMSA)}$ matrix can be written as a
sum of three factorized terms (as it can be immediately inferred by
expanding the square $\sigma_{ij}^2=(\sigma_i+\sigma_j)^2/4$) and has
the interesting physical interpretation of being proportional to the
area of the actual contact surface $4 \pi \sigma_{ij}^2$ between
particles of species $i$ and $j$. Finally, and for reasons related to the C1  
approximation that will be further elaborated below, we consider Case V 
defined by the linear (rather than quadratic) dependence
\bq \label{modelV}
\frac{1}{\tau_{ij}}=\frac{1}{\tau}\
\frac{\left\langle\sigma\right\rangle}{\sigma_{ij}}~, 
\eq
in this case the $K_{ij}^{(mMSA)}$ parameters can be written as a
sum of two factorized terms.

\section{The C1 approximation}
\label{sec:c1}
It was recently argued \cite{Gazzillo04} in the one-component case,
that the mMSA (C0) approximation can be improved by including the next
order term in the density expansion of the direct correlation
function. Its extention to multicomponent mixtures reads 
\begin{equation}
c_{ij}(r)=f_{ij}(r)[1+\sum_{m}\rho_m\gamma_{imj}^{(1)}(r)]~~~r\ge\sigma_{ij}~,
\label{c1_1}
\end{equation}
where 
\bq \nonumber
\gamma_{imj}^{(1)}(r)&=&\int f_{im}(|\rr-\rr^\prime|)f_{mj}(r^\prime)
\,d\rr^\prime \\ \label{c1:gamma1}
&=&\frac{2\pi}{r}\int_0^\infty ds\,sf_{im}(s)\int_{|r-s|}^{r+s}dt
\,tf_{mj}(t)~.
\eq 
is the first-order coefficient in the density expansion of the partial
indirect correlation functions $\gamma_{ij}(r)$. 
As discussed in Ref. \cite{Gazzillo04}, if we retain in the PY
closure only the terms corresponding to the zero and first-order
expansion in density we recover the C1 approximation (\ref{c1_1}). 
It turns out that Baxter factor
correlation function can still be cast in the form, 
Eqs. (\ref{m3})-(\ref{m5}) but the $K_{ij}$ parameters have the form
\bq
K^{(C1)}_{ij}=K^0_{ij}\,y^{(C1)}_{ij}(\sigma_{ij})~,
\eq
where the partial cavity functions at contact for this closure are
\bq \label{c1:y(sigma)}
y^{(C1)}_{ij}(\sigma_{ij})=1+\sum_{m}\rho_m\gamma^{(1)}_{imj}(\sigma_{ij})~,
\eq

Using in Eq. (\ref{c1:gamma1}) $f_{ij}(r)=-\theta(\sigma_{ij}-r)+
\delta(r-\sigma_{ij})\sigma_{ij}/(12\tau_{ij})$, we find after some
algebra the following result
\bq \nonumber
\gamma^{(1)}_{imj}(\sigma_{ij})&=&\frac{2\pi}{\sigma_{ij}}\left\{
\frac{\sigma_{im}^2}{12\tau_{im}}\left[-\frac{1}{2}(\sigma_{mj}^2-L_{jm}^2)
+\frac{\sigma_{mj}^2}{12\tau_{mj}}\right]+\right.\\ \nonumber
&&\frac{2}{3}\sigma_{ij}L_{mi}^3+\frac{\sigma_{mj}^2}{12\tau_{mj}}\frac{1}{2}
(L_{mi}^2-\sigma_{mi}^2)+\\ \nonumber
&&\frac{1}{4}(\sigma_{mj}^2-\sigma_{ij}^2)(\sigma_{mi}^2-L_{mi}^2)+
\frac{1}{3}\sigma_{ij}(\sigma_{mi}^3-L_{mi}^3)-\\ \label{c1-poly:gamma}
&&\left.\frac{1}{8}(\sigma_{mi}^4-L_{mi}^4)\right\}~,
\eq
Because of the presence of the factor $1/\sigma_{ij}$
in Eq. (\ref{c1-poly:gamma}), $K^{(C1)}_{ij}$ cannot be expressed as a
sum of factorized terms if we use any of the Cases I, II, or III. Case
IV, on the other hand, would be tractable, but it would yield
$K^{(C1)}_{ij}$ as a sum of 14 factorized
terms (proportional to $\sigma_i^n\sigma_j^m$ with $n, m=0,1,2,3$
except $n=m=0,3$) which is unmanageable in practice. In Case V, on the
other hand, a great simplification occurs and we find
\bq \label{c1-poly:Kij}
K^{(C1)}_{ij}=k_0+(\sigma_i+\sigma_j)k_1+\sigma_i\sigma_jk_2~,
\eq
where
\bq
k_0&=&\eta\frac{1}{576}
\frac{\langle\sigma\rangle^3\langle\sigma^2\rangle}{\langle\sigma^3\rangle}
\frac{1}{\tau^3}~,\\
k_1&=&\frac{1}{24}\langle\sigma\rangle\frac{1}{\tau}+\eta\left(
\frac{1}{576}\frac{\langle\sigma\rangle^4}{\langle\sigma^3\rangle}\frac{1}{\tau^3}
-\frac{1}{48}\frac{\langle\sigma\rangle^2\langle\sigma^2\rangle}{\langle\sigma^3\rangle}
\frac{1}{\tau^2}+\frac{1}{24}\langle\sigma\rangle\frac{1}{\tau}\right)~,\\
k_2&=&\eta\left(
\frac{1}{576}\frac{\langle\sigma\rangle^3}{\langle\sigma^3\rangle}\frac{1}{\tau^3}
-\frac{1}{24}\frac{\langle\sigma\rangle^3}{\langle\sigma^3\rangle}\frac{1}{\tau^2}
+\frac{1}{8}\frac{\langle\sigma\rangle\langle\sigma^2\rangle}{\langle\sigma^3\rangle}
\frac{1}{\tau}
\right)~,
\eq
where $\eta=\xi_3$ is the packing fraction. The expression (\ref{c1-poly:Kij})
is slightly more complicated than the $K^{(mMSA)}_{ij}$ treated with Case IV, 
because of the $k_0$ term. This noteworthy feature is the main justification for 
the particular form of Case V.
\section{Phase instabilities}
\label{sec:inst}
Our first task is the analysis of
the phase instabilities for the
polydisperse system only in the mMSA using Cases I, II, and III.

The next level of approximation (C1) is considerably more laborious
(since the calculations for the C1 approximation even in
the simple case of Case V requires determinants of $n$-dyadic
matrices with $n>4$) and we shall limit ourselves to the one-component
case for simplicity.

\subsection{mMSA approximation for the discrete polydisperse system}

For $p$-component mixtures, one can define the following generalization of
the Bhatia-Thornton concentration-concentration structure factor
\cite{Bhatia70,Gazzillo94,Gazzillo95} 
\begin{equation}
S_{{\rm CC}}\left( k\right) \ /\ \left( \prod_{m}x_{m}\right) =\left| {\bf S}
(k)\right| \ \sum_{i,j=1}^{p}\left( x_{i}x_{j}\right)
^{1/2}S_{ij}^{-1}\left( k\right)~,  \label{s1}
\end{equation}
where $\left|{\bf S}(k)\right| $ denotes the determinant of the matrix 
${\bf S}(k)$\ whose elements are the Ashcroft-Langreth partial structure factors
\cite{Ashcroft67}. 
Furthermore, the $S_{ij}^{-1}\left( k\right) $ functions are the elements of
the inverse of ${\bf S}(k)$, which can be expressed as 
\begin{equation}
S_{ij}^{-1}\left( k\right) =\delta _{ij}-\left( \rho _{i}\rho _{j}\right)
^{1/2}\widetilde{c}_{ij}\left( k\right) =\sum_{m}\widehat{Q}_{mi}\left(
-k\right) \widehat{Q}_{mj}\left( k\right)~,  \label{s2}
\end{equation}
with $\widetilde{c}_{ij}\left( k\right) $ three-dimensional Fourier
transform of $c_{ij}\left( r\right) $, $\widehat{Q}_{ij}\left( k\right)
=\delta _{ij}-2\pi (\rho _{i}\rho _{j})^{1/2}\widehat{q}_{ij}\left( k\right)
,$ and $\widehat{q}_{ij}\left( k\right) $ being the uni-dimensional Fourier
transform of $q_{ij}(r)$ ($k$ is the magnitude of the exchanged wave vector, 
$\delta _{ij}$ the Kronecker delta).

Phase instability corresponds to the divergence of the long wavelength limit 
$S_{{\rm CC}}\left( k=0\right) $, which is related to the concentration
fluctuations. Taking into account the relations
\begin{equation}
\sum_{i,j}\left( x_{i}x_{j}\right) ^{1/2}S_{ij}^{-1}\left( 0\right)
=\sum_{i}x_{i}a_{i}^{2}=\left( \rho k_{B}TK_{T}\right) ^{-1}=\left( 
\frac{\partial \beta P}{\partial \rho }\right) _{T}~,  \label{s3}
\end{equation}
\begin{equation}
\left| {\bf S}(0)\right| =\left| {\bf I}-{\bf C}(0)\right| ^{-1}=\left| 
\widehat{{\bf Q}}(0)\right| ^{-2}~,  \label{s4}
\end{equation}
[where $K_{T}$ is the isothermal
compressibility, ${\bf I}$ the unit matrix of order $p$, and ${\bf C}$ has
elements $(\rho _{i}\rho _{j})^{1/2}\widetilde{c}_{ij}\left( k\right)
$], $S_{{\rm CC}}\left( k=0\right)$ can be re-expressed as 
\begin{equation}
\frac{S_{{\rm CC}}\left( 0\right) }{\prod_{m}x_{m}}=\frac{1}{\left| 
\widehat{{\bf Q}}(0)\right| ^{2}\left( \rho k_{B}TK_{T}\right) }~.  \label{s5}
\end{equation}

For a one-component system the divergence of $K_{T}$ signals
mechanical instability, associated with a 
gas-liquid phase transition or condensation. However, a multi-component
fluid usually becomes unstable while $K_{T}$ remains finite and different
from zero. In this case, it is the vanishing of 
$\left| \widehat{{\bf Q}}(0)\right|$ which causes the divergence of 
$S_{{\rm CC}}\left(0\right)$ and produces a phase instability
\cite{Gazzillo94,Gazzillo95}. Indeed if one tries to calculate the
locus of points in the phase diagram $(\tau,\eta)$ where
$\sum_ix_ia_i^2=0$, using Cases I, II, or III, discovers that such
curves disappear (the quadratic equations in $\tau$ have a negative
discriminant) as soon as we switch on the size polydispersity letting
$\langle\sigma^2\rangle\ne\langle\sigma\rangle^2$. We remark that the
exact nature of this instability requires a more involved analysis and
it will be deferred to a future work.

The computation of $\left| \widehat{{\bf Q}}(0)\right| ,$ which usually
becomes a formidable task with increasing $p,$ is rather simple for the mMSA
solution of Baxter model when $K_{ij}$ is factorized as in
Eq. (\ref{m10}). In fact, $\widehat{{\bf Q}}(k)$ becomes a $n$-{\sl
dyadic} (or Jacobi) matrix 
\begin{equation}
\widehat{Q}_{ij}=\delta _{ij}+\sum_{\nu =1}^{n}A_{i}^{(\nu )}B_{j}^{(\nu
)}\qquad (i,j=1,\ldots ,p)~,  \label{s6}
\end{equation}
with the remarkable property that its determinant, which is of order $p$,
turns out to be equal to a determinant of order $n$ ($\ll p$ for
polydisperse fluids) \cite{Gazzillo00}. The necessary expressions are
reported in Appendix \ref{app:dyadic}.

For factorized $K_{ij}$'s, one finds 
\begin{equation}
\widehat{Q}_{ij}\left( 0\right) =\delta _{ij}+\frac{\pi }{6}\left( \rho
_{i}\rho _{j}\right) ^{1/2}\left[ \frac{1}{\Delta }\sigma _{j}^{3}+\sigma
_{i}\frac{3}{\Delta }\left( \xi _{2}\frac{1}{\Delta }\sigma _{j}^{3}+\sigma
_{j}^{2}\right) -12Y_{i}\left( \xi _{1,1}\frac{1}{\Delta }\sigma
_{j}^{3}+\sigma _{j}Y_{j}\right) \right]~,  \label{s8}
\end{equation}
with
\begin{equation}
\xi _{m,n}=\frac{\pi }{6}\rho \left\langle \sigma ^{m}Y^{n}\right\rangle~,
\label{s9}
\end{equation}
($\left\langle \cdots \right\rangle $ denotes a compositional average, i.e. $%
\left\langle FG\right\rangle \equiv \sum_{i}x_{i}F_{i}G_{i}$). Note that $%
\xi _{m,0}=\xi _{m}.$

We emphasize that the decomposition of Eq. (\ref{s8}) into $A_{i}^{(\nu )}$
and $B_{j}^{(\nu )}$ is not unique. However, $\widehat{Q}_{ij}\left(
0\right)$ of Case I and III is 3-dyadic (i.e. it contains $n=3$ dyadic terms),
while $\widehat{Q}_{ij}\left( 0\right) $ of Case II is simply 2-dyadic. As
a consequence, one has to calculate at most a determinant of order 3. The
general result for all three cases is 
\begin{equation}
\left| \widehat{{\bf Q}}(0)\right| =\frac{1}{\Delta ^{2}}\left[ \left(
1+2\xi _{3}\right) \left( 1-12\xi _{1,2}\right) +36\xi _{2,1}^{2}\right]~.
\label{s10}
\end{equation}

Physically admissible states must satisfy the inequality
$\left|\widehat{{\bf Q}}(0)\right|>0$ \cite{Barboy74} and
the stability boundary is reached when 
$\left|\widehat{{\bf Q}}(0)\right|=0$, which yields
\begin{equation}
\label{s11}
\tau =\left\{
\begin{array}{ll}
\displaystyle
\frac{\left\langle \sigma \right\rangle ^{3}}{\left\langle \sigma
^{3}\right\rangle }\eta -\left( \frac{\left\langle \sigma \right\rangle
\left\langle \sigma ^{2}\right\rangle }{\left\langle \sigma
^{3}\right\rangle }\right) ^{2}\frac{3\eta ^{2}}{1+2\eta } &
\text{Case I}~,\\
\displaystyle 
\frac{\eta \left( 1-\eta \right) }{1+2\eta } & \text{Case II}~,\\
\displaystyle
\frac{\langle\sigma\rangle\langle\sigma^2\rangle}{\langle\sigma^3\rangle}
\eta-\frac{\langle\sigma^2\rangle^3}{\langle\sigma^3\rangle^2}
\frac{3\eta^2}{1+2\eta} & \text{Case III}~.
\end{array}\right.
\end{equation}
If the HS diameters follow a Schulz distribution, then the stability
boundary of Cases I and III can be expressed as
\bq \label{s13}
\tau =\left\{
\begin{array}{ll}
\displaystyle
\eta \left(\frac{1}{M_1M_2}-\frac{1}{M_2^2}\frac{3\eta }{1+2\eta }\right) &
\text{Case I}~,\\
\displaystyle
\eta \left(\frac{1}{M_2}-\frac{M_1}{M_2^2}\frac{3\eta }{1+2\eta }\right) &
\text{Case III}~,
\end{array}\right.
\eq
where $M_j=1+js^2$
with $s=\left[ \left\langle \sigma ^{2}\right\rangle -\left\langle \sigma
\right\rangle ^{2}\right] ^{1/2}/\left\langle \sigma \right\rangle $
measuring the degree of size polydispersity.

The fluid is stable at `temperatures' $\tau$ higher than those given
by the previous equations (since $|\widehat{{\bf Q}}(0)|>0$). Let us
now compare two mixtures with the same packing fraction $\eta$ but
different polydispersity degree $s$. As depicted in
Fig. \ref{fig:mech} at small $\eta$ values, increasing $s$ at fixed
$\eta$ lowers the stability curve of Case I and III.
As shown by the left branch of the curve (the opposite trend on the
right hand side of the figure is not acceptable, 
since the mMSA closure can be a reasonable approximation only in the low
density regime) the onset of instability occurs at lower $\tau $. 
As expected, polydispersity renders the mixture more
stable with respect to concentration fluctuations. Quite surprisingly, 
on the other hand, the stability boundary does not depend on $s$ 
at fixed $\eta $ in Case II, and all 
mixtures with different polydispersity have the same stability boundary as
the one-component case ($s=0$).

\subsection{C1 approximation for the one component system}

As remarked, the C1 approximation yields rather more complex expressions
and here we restrict to the one-component case. Yet, this example provides
a flavor of how this approximation would work in the multicomponent
case and could be compared with the result given by 
$\left|\widehat{{\bf Q}}(0)\right|=0$.
For the one-component system phase instability coincides with the
divergence of $K_T$. As from Eq. (\ref{s3})
\bq \label{C1:mech}
(\rho k_BTK_T)^{-1}=a^2=\left[\frac{1+2\eta}{(1-\eta)^2}-\frac{1}{\tau}
y^{(C1)}(\sigma)\frac{\eta}{1-\eta}\right]=0~,
\eq
where [see Eqs. (\ref{c1:y(sigma)}) and (\ref{c1-poly:gamma})]
\bq
y^{(C1)}(\sigma)=1+y_1(\tau)\eta~,
\eq
with
\bq
y_1(\tau)=\frac{5}{2}-\frac{1}{\tau}+\frac{1}{12\tau^2}~.
\eq

The curve for the onset of mechanical instability is shown in
Fig. \ref{fig:mech} and compared with the PY one
\bq \label{PY:mech}
\tau=\frac{10-9/(1-\eta)+14\eta}{12(1+2\eta)}~.
\eq
One clearly sees that the C1 stability boundary lowers and shifts to
the left in agreement with the PY result. 

\section{Percolation threshold}
\label{sec:perc}
In view of the simplicity of the mMSA (C0) solution, one might expect
that other quantities, besides those discussed so far, 
can be computed analytically. We now show that this is indeed the case.
The problem we address in this section is {\sl continuum percolation}.
This problem is far from being new \cite{Hill55}. However new
activity along this line has been stirred by recent and 
precise Monte Carlo results for the one-component case
\cite{Miller03,Miller04}, and it is then rather 
interesting to consider its multicomponent extension.
For the sake of completeness we now recall the basic necessary formalism
\cite{Coniglio77,Chiew83,Chiew89}. 

In the sticky limit the partial Boltzmann factors read
\bq
e_{ij}(r)=\theta(r-\sigma_{ij})+\frac{K^0_{ij}}{\sigma_{ij}}
\delta(r-\sigma_{ij})~,  
\eq
where $\theta$ is the Heaviside step function and $\delta$ the Dirac
delta function.

When studying percolation problems in the continuum is useful to
rewrite the Boltzmann factor as the sum of two terms
\cite{Hill55,Coniglio77} $e_{ij}(r)=e^*_{ij}(r)+e^+_{ij}(r)$, where
\bq
e^*_{ij}(r)&=&\theta(r-\sigma_{ij})~,\\
e^+_{ij}(r)&=&\frac{K^0_{ij}}{\sigma_{ij}}
\delta(r-\sigma_{ij})~.
\eq
The corresponding Mayer functions will be
$f_{ij}(r)=f^*_{ij}(r)+f^+_{ij}(r)$, with
\bq
f^*_{ij}(r)&=&e^*_{ij}(r)-1~,\\
f^+_{ij}(r)&=&e^+_{ij}(r)~.
\eq

The procedure to obtain equations of {\sl connectedness} and {\sl blocking}
functions from the usual pair correlation functions and direct
correlation functions is best described through the use of graphical
language. If we substitute $f^*_{ij}$ and $f^+_{ij}$ bonds for
$f_{ij}$ bonds in the density expansions for these functions, then the
connectedness functions, which we will indicate with a cross
superscript, are expressed as the sums of those terms that have at
least one $f^+_{ij}$ bond path connecting the two root vertexes. The
sums of the remaining terms in the expansions give the blocking
functions. 

The percolation threshold corresponds to the existence of an
infinite cluster of particles and is given by the divergence of the mean
cluster size \cite{Hill55,Coniglio77}
\bq \nonumber
S_{cluster}&=&1+\rho\sum_{i,j}x_ix_j\int d\rr\,h^+_{ij}(r)\\
&=& S^+_{NN}(k=0)\equiv\sum_{i,j}(x_ix_j)^{1/2}S^+_{ij}(k=0)~,
\eq
where $h^+_{ij}(r)$ is the pair connectedness function (related to the
joint probability of finding a particle of species $i$ and a particle
of species $j$ at a distance $r$ and that these two particles are
connected) and
\bq
S^+_{ij}(k)\equiv\delta_{ij}+(\rho_i\rho_j)^{1/2}\,\tilde{h}^+_{ij}(k)~.
\eq

Since $h^+_{ij}(r)$ is related to the so called direct connectedness
function $c^+_{ij}(r)$ through an OZ equation, one can use Baxter
formalism again, introducing a factor function $q^+_{ij}(r)$. If we
now define $\widehat{Q}_{+,ij}(k)=\delta_{ij}-2\pi(\rho_i\rho_j)^{1/2}
\,\widehat{q}^+_{ij}(k)$, then it results that
\bq
S^+_{ij}(k)=\sum_m\widehat{Q}^{-1}_{+,im}(k)\widehat{Q}^{-1}_{+,jm}(-k)~,
\eq  
and thus
\bq\label{mMSA:Scluster}
S_{cluster}=\sum_m s^2_m(0)~,
\eq
where 
\bq\label{mMSA:sm}
s_m(0)=\sum_i\sqrt{x_i}\,\widehat{Q}^{-1}_{+,im}(0)~.
\eq
Clearly $\widehat{Q}^{-1}_{+,im}(0)$ diverges to infinity when
$|\widehat{\QQ}_+(0)|=0$, and this relation defines the percolation
threshold.  

Another interesting and related quantity is the average coordination
number 
\bq \label{barZ}
\bar{Z}&=&4\pi\rho\sum_{i,j}x_ix_j\int_0^{\sigma_{ij}}h^+_{ij}(r)r^2\,dr~.
\eq

\subsection{mMSA approximation}

The mMSA closure for $c^+_{ij}(r)$ is 
\bq\label{mMSA:c+closure}
c^+_{ij}(r)=f^+_{ij}(r)=0~~~r>\sigma_{ij}~,
\eq

On the other hand when $r\le \sigma_{ij}$ we have $e^*_{ij}(r)=0$ and
$f^+_{ij}(r)=e_{ij}(r)$, so we must have exactly
\bq\nonumber
h^+_{ij}(r)&=&e^*_{ij}(r)y^+_{ij}(r)+f^+_{ij}(r)y_{ij}(r)\\\nonumber
&=&e_{ij}(r)y_{ij}(r)\\ \label{mMSA:h+closure}
&=&\frac{K^0_{ij}}{\sigma_{ij}}y_{ij}(\sigma_{ij})\delta(r-\sigma_{ij})~~~
r\le\sigma_{ij}~.
\eq

Within the mMSA we have for the cavity function at contact
\cite{Gazzillo04} 
\bq
y_{ij}(\sigma_{ij})=1~~~\mbox{for all $i,j$}~.
\eq
Following the same steps of Chiew and Glandt \cite{Chiew83,Chiew89} we then
find (see Appendix \ref{app:q+} for details) 
\bq\label{mMSA:q+}
q^+_{ij}(r)=K_{ij}\theta(r-L_{ij})\theta(\sigma_{ij}-r)~.
\eq
From which it follows
\bq \label{perc:Q+}
\widehat{Q}_{+,ij}(0)=\delta_{ij}-2\pi(\rho_i\rho_j)^{1/2}K_{ij}\sigma_j~.
\eq
Within Cases I, II, and III
\bq \label{mMSA:Q+}
\widehat{Q}_{+,ij}(0)&=&\delta_{ij}+a_i^+b_j^+~,\\
a_i^+&=&-2\pi\rho\sqrt{x_i}\,Y_i~,\\
b_j^+&=&\sqrt{x_j}\,Y_j\sigma_j
\eq

Now from Eq. (\ref{mMSA:Q+}) follows that $\widehat{Q}_{+,ij}(0)$ is a
1-dyadic form. Using the properties of dyadic matrices (see Appendix
\ref{app:dyadic}) we then find  
\bq
\widehat{Q}^{-1}_{+,ij}(0)=\frac{1}{|\widehat{\QQ}_+(0)|}
\left|
\begin{array}{cc}
\delta_{ij} & b_j^+\\
a_i^+ & 1+{\mathbf a^+}\cdot{\mathbf b^+}
\end{array}
\right|~,
\eq
where 
\bq
|\widehat{\QQ}_+(0)|=1+{\mathbf a^+}\cdot{\mathbf b^+}=1-12\xi_{1,2}~.
\eq
From Eq. (\ref{mMSA:sm}) we find
\bq
s_m(0)=\frac{1}{|\widehat{\QQ}_+(0)|}\left[\sqrt{x_m}\,(1+{\mathbf a^+}
\cdot{\mathbf b^+})-b_m^+\sum_i\sqrt{x_i}\,a_i^+\right]~,
\eq
and from Eq. (\ref{mMSA:Scluster})
\bq
S_{cluster}=1+\frac{24}{\xi_0}\frac{\xi_{1,1}\xi_{0,1}}{1-12\xi_{1,2}}
+\frac{144}{\xi_0}\frac{\xi_{2,2}\xi_{0,1}^2}{(1-12\xi_{1,2})^2}~.
\eq
The percolation transition occurs when
\bq \label{mMSA:percolation}
\tau=\left\{
\begin{array}{ll}
\displaystyle \frac{\langle\sigma\rangle^3}{\langle\sigma^3\rangle}\,\eta=
\frac{1}{M_1M_2}\,\eta & \mbox{Case I}~,\\
\displaystyle \eta & \mbox{Case II}~,\\
\displaystyle
\frac{\langle\sigma\rangle\langle\sigma^2\rangle}{\langle\sigma^3\rangle}
\,\eta=\frac{1}{M_2}\,\eta & \mbox{Case III}~.
\end{array}
\right.
\eq

The threshold is independent of $s$ at fixed $\eta$ for Case II, but
lowers with increasing size polydispersity in Cases I and III. The
curve is simply a straight line, as a consequence of the mean-field
character of the mMSA (C0) closure. The qualitative result found with
Cases I and III is however interesting.
For the average coordination number we find from Eqs. (\ref{barZ}) and
(\ref{mMSA:h+closure}) 
\bq\nonumber
\bar{Z}&=&4\pi\rho\sum_{i,j}x_ix_jK_{ij}\sigma_{ij}\\ 
&=&\frac{24}{\xi_0}\xi_{1,1}\xi_{0,1}\\\nonumber
&=&\left\{
\begin{array}{ll}
\displaystyle 2\frac{\eta}{\tau}\frac{\langle\sigma\rangle^3}
{\langle\sigma^3\rangle} & \mbox{Case I}~,\\
\displaystyle 2\frac{\eta}{\tau}\frac{\langle\sigma\rangle
\langle\sigma^2\rangle}{\langle\sigma^3\rangle} & \mbox{Case II,III}~.
\end{array}
\right.
\eq 
At the percolation transition we then find
\bq
\bar{Z}=\left\{
\begin{array}{ll}
2 & \mbox{Case I,III}~,\\
2/M_2 & \mbox{Case II}~.
\end{array}
\right.
\eq

Using Case IV $\widehat{Q}_{+ij}(0)$ turns out to be 3-dyadic; the
percolation transition occurs when $|\widehat{\QQ}_+(0)|=0$, i.e.
\bq \label{mMSA:perc-modelIV}
1-\frac{\eta}{\tau}-\frac{s^2(4+7s^2)}{8(1+3s^2+2s^4)}\left(\frac{\eta}{\tau}\right)^2
+\frac{s^6}{16(1+s^2)(1+2s^2)^2}\left(\frac{\eta}{\tau}\right)^3=0~.
\eq
The solution $\eta/\tau=p(s)$ such that $p(0)=1$ is a monotonously
decreasing function with 
\bq
\lim_{s\to\infty}p(s)=0.756431\ldots~.
\eq 
Then with this Case we find that increasing the polydispersity the
non-percolating region of the phase diagram diminishes.

With Case V $\widehat{Q}_{+ij}(0)$ turns out to be 2-dyadic, and the
percolation transition occurs when 
\bq \nonumber
\tau&=&\left(
\frac{\langle\sigma\rangle\langle\sigma^2\rangle}{\langle\sigma^3\rangle}+
\sqrt{\frac{\langle\sigma\rangle^3}{\langle\sigma^3\rangle}}\right)
\frac{\eta}{2}\\
\label{mMSA:perc-modelV}
&=&\left(\frac{1}{M_2}+\sqrt{\frac{1}{M_1M_2}}\right)\frac{\eta}{2}~,
\eq
which has the physical behavior already found with Cases I, II, and
III.

\subsection{C1 approximation with Case V}
\label{sec:C1percolation}
As remarked, in Case V we can work out the percolation threshold equation
even within the C1 approximation. From Eq. (\ref{mMSA:h+closure}) 
we have exactly 
\bq
h^+_{ij}(r)=\frac{K^0_{ij}}{\sigma_{ij}}y^{(C1)}_{ij}(\sigma_{ij})\delta(r-\sigma_{ij})
~~~r\le\sigma_{ij}~.
\eq
where $y_{ij}^{(C1)}(\sigma_{ij})$ is given by Eqs. (\ref{c1:y(sigma)}). 
For the closure condition of the direct connectedness function we find again
\bq\nonumber
c^+_{ij}(r)&=&f^+_{ij}(r)+f^+_{ij}(r)\sum_m\rho_m\gamma^{(1)}_{imj}(r)+f^*_{ij}(r)
\sum_m\rho_m\gamma^{(1)+}_{ijm}(r)\\
&=&0~~~r>\sigma_{ij}~,
\eq
since $f^+_{ij}(r)=f^*_{ij}(r)=0$ for $r>\sigma_{ij}$. To determine
$q^+_{ij}(r)$ we then follow the same steps as for the mMSA case and
we find 
\bq
q^+_{ij}(r)=K^0_{ij}y^{(C1)}_{ij}(\sigma_{ij})\theta(r-L_{ij})
\theta(\sigma_{ij}-r)~.   
\eq

When we insert $K_{ij}$ from Eq. (\ref{c1-poly:Kij}) into the expression for
$\widehat{Q}_{+ij}(0)$ [see Eq. (\ref{perc:Q+})] this becomes a 4-dyadic matrix
whose determinant is  
\bq\label{C1:perctrans}
|\widehat{\bf Q}_+(0)|=1+\sum_{i=1}^6 q_i(s,\eta)/\tau^i~,
\eq
where the coefficients $q_i(s,\eta)$ are given in Appendix \ref{app:c1-perc}.

The percolation threshold is the solution of $|\widehat{\bf
Q}_+(0)|=0$. This is an algebraic equation of order 6 in $\tau$. 
We can plot the correct root $\tau(\eta)$ for different values of
polydispersity, as reported in Fig. \ref{fig:percolation-c1-poly}.
We see that increasing the polydispersity increases
the non-percolating phase. One can clearly observe a clear improvement
from the mMSA (C0) approximation although the $\eta \to 0$ limit is
still qualitatively different from the PY one-component case. It would be
interesting to study if the ``true'' percolation threshold passes
through the origin $(\eta=0,\tau=0)$ (as occur in the C0 or C1
approximations) or has a finite limit $(\eta=0,\tau=\tau_0)$ 
(as it occur for monodisperse fluids in the PY approximation with
$\tau_0=1/12$). Even if the 
Monte Carlo results of Ref. \cite{Miller03,Miller04} are inconclusive
in this respect, physically it
is plausible to assume that at very low density the average number of
bonds per particle is not sufficient to support large clusters at all
and we would tend to favour the first scenario
\footnote{In this respect both C0 and C1 would be more precise than
the PY closure and this is a remarkable feature.}.

For the one-component system the average cluster size is
\bq\nonumber
S_{cluster}&=&1+\rho\tilde{h}^+(0)=\frac{1}{1-\rho\tilde{c}^+(0)}=
\frac{1}{[\widehat{Q}_+(0)]^2}\\
&=&\frac{1}{[1-\eta y^{(C1)}(\sigma)/\tau]^2}~.
\eq
The percolation transition occurs when $\eta y^{(C1)}(\sigma)=\tau$ or
\bq \label{C1:percolation}
\eta=\frac{2\left(-3\tau^2+\sqrt{3}\tau^{3/2}\sqrt{1-9\tau+30\tau^2}\right)}
{1-12\tau+30\tau^2}~.
\eq
In Fig. \ref{fig:binodal-c1} we compare our result for the
one-component ($s=0$) system with the PY result of Chiew
and Glandt \cite{Chiew83} and the Monte Carlo simulation of Miller and
Frenkel \cite{Miller03,Miller04}.
 
The average coordination number becomes 
\bq
\bar{Z}=2\frac{\eta}{\tau}y^{(C1)}(\sigma)~,
\eq
and at the percolation transition we find $\bar{Z}=2$.

\section{Phase equilibrium}
\label{sec:binodal}

Phase equilibrium is another interesting aspect which can be analyzed 
in full details within our model.
It was pointed out in Ref.
\cite{Gazzillo03} that the equation of state derived from the energy
route for a one-component system of sticky hard spheres in the mMSA
approximation is van der Waals like. The same holds true for the system
studied with the C1 approximation. It is worth stressing that 
the equation of state derived from
the compressibility route cannot yield a van der Waals loop since
from Eq. (\ref{s3}) $[\partial (\beta P)/\partial\rho]_T>0$ 
\footnote{Even though it may happen that one has loss of solution of
$\sum_ix_ia_i^2$ for certain values of the density, as occurs for the
Percus Yevick closure \cite{Baxter68}.}. 
On the other hand the equation of state derived from the virial equation has
been shown to diverge for the mMSA approximation \cite{Gazzillo04} and
we anticipate that it also diverges for the C1 approximation.
This is the reason why we focus our analysis on the energy route in
the present work.

In this section we will find the binodal curves for the polydisperse
system treated with the mMSA (C0) approximation and for the one-component
system treated with the C1 approximation. The coexistence problem for
a polydisperse system is, in general, a much harder task than its 
one-component counterpart, since it involves the solution of a large 
(infinite) number of integral non-linear equations.  
But we will see that since our excess free energy is
expressed in terms of a finite number of moments of the size
distribution function (a similar feature occurs for polydisperse van 
der Waals models \cite{Bellier00}, for polydisperse HS \cite{Sollich02} 
and for Yukawa-like potentials \cite{Kalyuzhnyi03,Kalyuzhnyi04}) 
the coexistence problem can be simplified and
becomes numerically solvable through a simple Newton-Raphson
algorithm [see Eq. (\ref{tp:eq1})-(\ref{tp:eq3})].
The necessary formalism to this aim can be found in a recent
review \cite{Sollich02}, and we will briefly recall it next.    

\subsection{From a discrete to a continuous polydisperse mixture}

Consider a mixture made of $p$ components labeled $i=1,\ldots,p$,
containing $N^{(0)}$ particles and with density $\rho^{(0)}$, which
separates, at a certain temperature $\tau$, into $m$ daughter phases,
where each phase, labeled $\alpha=1,\ldots,m$, has a number of particles
$N^{(\alpha)}$ and density $\rho^{(\alpha)}$. Let the molar fraction
of the particles of species $i$ of phase $\alpha$ be
$x_i^{(\alpha)}$, $\alpha=0$ corresponding to the parent phase. At
equilibrium the following set of constraints must 
be fulfilled: (i) volume conservation, (ii) conservation of the total
number of particles of each species, (iii) equilibrium condition for the
pressures $P^{(\alpha)}(\tau,\rho^{(\alpha)},\{x_i^{(\alpha)}\})$, and
(iv) equilibrium condition for the chemical potentials
$\mu_i^{(\alpha)}(\tau,\rho^{(\alpha)},\{x_i^{(\alpha)}\})$. This set
of constraints form a closed set of equations (see Appendix
\ref{app:coex} for details) for the $(2+p)m$ unknowns
$\rho^{(\alpha)}$, $x^{(\alpha)}=N^{(\alpha)}/N^{(0)}$, and
$x_i^{(\alpha)}$ with $i=1,\ldots,p$ and
$\alpha=1,\ldots,m$. Extension to the polydisperse case with an
infinite number of components is achieved by switching from the
discrete index variable $i$ to the continuous variable $\sigma$ using
the following {\sl replacement rule} 
\bq \label{polycoex:dist}
x_i\to F(\sigma)d\sigma~,
\eq
where $F(\sigma)d\sigma$ is the fraction of particles with diameter in
the interval $(\sigma,\sigma+d\sigma)$. The function $F(\sigma)$ will
be called {\sl molar fraction density function} or more simply size
distribution function. Notice that, due to this replacement rule, we
also have
\bq \label{polycoex:P}
P^{(\alpha)}(\tau,\rho^{(\alpha)},\{x_i^{(\alpha)}\})&\to&
P^{(\alpha)}(\tau,\rho^{(\alpha)};[F^{(\alpha)}])~,\\
\label{polycoex:mu}
\mu_i^{(\alpha)}(\tau,\rho^{(\alpha)},\{x_i^{(\alpha)}\})&\to&
\mu^{(\alpha)}(\sigma,\tau,\rho^{(\alpha)};[F^{(\alpha)}])~,
\eq
i.e. the thermodynamic quantities become functionals of the size
distribution function and the equilibrium conditions (ii)-(iv) has to be
satisfied for all values of the continuous variable $\sigma$. The
phase coexistence problem that now consists in solving the constraints
(i)-(iv) for the unknowns $\rho^{(\alpha)}$, $x^{(\alpha)}$, and
$F^{(\alpha)}(\sigma)$ for $\alpha=1,\ldots,m$, turns out to be a
rather formidable task hardly solvable from a numerical point of view.
Fortunately, as outlined in the next subsection, for our model a
remarkable simplification occurs. 

\subsection{Truncatable excess free energy}

As is described in the next subsection, the excess free energy of our
system is {\sl truncatable}: it is only a function of the three
moments $\xi_i$, $i=1,2,3$ of the size distribution function [see
Eq. (\ref{mMSA:bA/N}) for Case I, II, III, IV, and V
treated with mMSA, and Eq. (\ref{C1:bA/N}) for Case V treated with
C1]. So we have the following simplification
\bq
P^{(\alpha)}(\tau,\rho^{(\alpha)};[F^{(\alpha)}])&\to&
P^{(\alpha)}(\tau,\rho^{(\alpha)};\{\xi_i^{(\alpha)}\})~,\\
\mu^{(\alpha)}(\sigma,\tau,\rho^{(\alpha)};[F^{(\alpha)}])&\to&
\mu^{(\alpha)}(\sigma,\tau,\rho^{(\alpha)};\{\xi_i^{(\alpha)}\})~,
\eq
where $\{\xi_i^{(\alpha)}\}$ is a short-hand notation for 
$\xi_1^{(\alpha)},\xi_2^{(\alpha)},\xi_3^{(\alpha)}$.
It turns out that the two-phase ($m=2$) coexistence problem, the one
in which we are interested (we are thus concentrating on the high
temperature portion of the phase diagram), reduces to the solution of
the following eight equations in the eight unknowns $\rho^{(1)}$,
$\rho^{(2)}$, $\{\xi_i^{(1)}\}$, and $\{\xi_i^{(2)}\}$
\bq \nonumber 
\xi_i^{(\alpha)}&=&\frac{\pi}{6}\rho^{(\alpha)}\int 
Q^{(\alpha)}(\sigma,\tau,\rho^{(0)},\rho^{(1)},\rho^{(2)};
\{\xi_i^{(1)}\},\{\xi_i^{(2)}\})F^{(0)}(\sigma)\sigma^i\,d\sigma~,\\
\label{tp:eq1}
&&i=1,2,3~~~\alpha=1,2~,\\ \nonumber
1&=&\int Q^{(\alpha)}(\sigma,\tau,\rho^{(0)},\rho^{(1)},\rho^{(2)};
\{\xi_i^{(1)}\},\{\xi_i^{(2)}\})F^{(0)}(\sigma)\,d\sigma~,\\
\label{tp:eq2}
&&\alpha=\mbox{1 or 2}~,
\eq
\bq \label{tp:eq3}
P^{(1)}(\tau,\rho^{(1)};\{\xi_i^{(1)}\})=
P^{(2)}(\tau,\rho^{(2)};\{\xi_i^{(2)}\})~,
\eq
with
\bq \label{tp:Q}
\rho^{(\alpha)}Q^{(\alpha)}=\rho^{(0)}
\frac{(\rho^{(1)}-\rho^{(2)})(1-\delta_{1\alpha}+\delta_{1\alpha}
e^{\beta\Delta\mu^{exc}})}{(\rho^{(1)}-\rho^{(0)})+
(\rho^{(0)}-\rho^{(2)})e^{\beta\Delta\mu^{exc}}}~,
\eq
and
\bq
\Delta\mu^{exc}&=&{\mu^{exc}}^{(2)}(\sigma,\tau,\rho^{(2)};[F^{(2)}])-
{\mu^{exc}}^{(1)}(\sigma,\tau,\rho^{(1)};[F^{(1)}])~,
\eq
where we indicate with the superscript $exc$ the excess part (over the
ideal) of the chemical potential.
For a complete derivation of Eqs. (\ref{tp:eq1})-(\ref{tp:eq3}) see
Appendix \ref{app:coex}. 
 
\subsection{Thermodynamic properties}

In order to obtain the equation of state of our model Eq. (\ref{m1}) from
the energy route, one exploits the following exact result \cite{Watts71}
\bq\nonumber
\frac{\partial(\beta A^{exc}/N)}{\partial\tau}&=&2\pi\rho\sum_{i,j}x_ix_j
\int\frac{\partial[\beta\phi_{ij}(r)]}{\partial\tau}\,g_{ij}(r)r^2dr\\ 
\nonumber
&=&2\pi\rho\sum_{i,j}x_ix_j\int_{\sigma_{ij}}^{R_{ij}}\frac{1}{\tau}\,
e_{ij}(r)y_{ij}(r)r^2dr\\\nonumber
&=&2\pi\rho\sum_{i,j}x_ix_j\frac{1}{\tau}\int_{\sigma_{ij}}^{R_{ij}}
\frac{1}{12\tau_{ij}}\frac{R_{ij}}{R_{ij}-\sigma_{ij}}y_{ij}(r)r^2dr~.
\eq
Upon taking the sticky limit we find
\bq \label{dadtau}
\frac{\partial(\beta A^{exc}/N)}{\partial\tau}&=&\frac{\eta}
{\langle\sigma^3\rangle}\frac{1}{\tau}\sum_{i,j}x_ix_j
\frac{1}{\tau_{ij}}\,\sigma_{ij}^3y_{ij}(\sigma_{ij})~.  
\eq

\subsubsection{mMSA approximation}

Within the mMSA approximation the partial cavity functions at contact are
all equal to 1 so from Eq. (\ref{dadtau}), after integration over
$\tau$ from $\tau=\infty$ (hard sphere case), we find 
\bq \label{mMSA:bA/N}
\frac{\beta(A^{exc}_{SHS}-A^{exc}_{HS})}{N}\xi_0=\left\{
\begin{array}{ll}
\displaystyle
-\frac{1}{\tau}\frac{\xi_1^3}{\xi_0} & \mbox{Case I}~,\\
\displaystyle
-\frac{1}{\tau}\xi_2\xi_1 &  \mbox{Case II, III}~,\\
\displaystyle
-\frac{1}{\tau}\frac{1}{4}\left(3\xi_1\xi_2+
\xi_0\xi_3\right) & \mbox{Case IV}~,\\ 
\displaystyle
-\frac{1}{\tau}\frac{1}{2}\left(\xi_1\xi_2+
\frac{\xi_1^3}{\xi_0}\right) & \mbox{Case V}~. 
\displaystyle 
\end{array}
\right.
\eq

The pressure can be found, from $\beta P/\rho=\eta\partial(\beta A/N)/
\partial\eta$
\bq \label{mMSA:Z}
\frac{\pi}{6}\beta[P_{SHS}(\tau,\rho;\{\xi_i\})-
P_{HS}(\tau,\rho;\{\xi_i\})]=
\frac{\beta(A^{exc}_{SHS}-A^{exc}_{HS})}{N}\xi_0~,
\eq
where
for $P_{HS}$ we use an equation due to Boubl\'ik, Mansoori, Carnahan,
Starling, and Leland (BMCSL) \cite{Boublik70,Mansoori71} 
which reduces to the Carnahan-Starling one when $s=0$,
\bq \label{polycoex:PHS}
\frac{\pi}{6}\beta
P_{HS}(\tau,\rho;\{\xi_i\})&=&Z_{HS}\xi_0=\frac{\xi_0}{1-\xi_3}+
3\frac{\xi_1\xi_2}{(1-\xi_3)^2}+3\frac{\xi_2^3}{(1-\xi_3)^3}-
\frac{\xi_3\xi_2^3}{(1-\xi_3)^3}\\ \nonumber
&=&\xi_0\left\{\frac{1}{1-\eta}+\frac{3\eta}{(1-\eta)^2}\frac{1}{M_2}+
\left[\frac{3\eta^2}{(1-\eta)^3}-\frac{\eta^3}{(1-\eta)^3}\right]
\frac{M_1}{M_2^2}\right\}~. 
\eq

The excess free energy of the polydisperse hard sphere system is
obtained integrating $(Z_{HS}-1)/\eta$ over $\eta$, from $\eta=0$, 
and recalling that the excess free energy is zero when $\eta=0$. We
then find \cite{Salacuse82}
\bq \nonumber
\frac{\beta A^{exc}_{HS}}{N}&=&\frac{\eta}{(1-\eta)^2}\frac{M_1}{M_2^2}+
\frac{3\eta}{1-\eta}\frac{1}{M_2}+\left[\frac{M_1}{M_2^2}-1\right]
\ln(1-\eta)\\ \label{BMCSL:A}
&=&\frac{\xi_2^3}{\xi_0\xi_3(1-\xi_3)^2}+
3\frac{\xi_1\xi_2}{\xi_0(1-\xi_3)}+
\left(\frac{\xi_2^3}{\xi_0\xi_3^2}-1\right)\ln(1-\xi_3)~.
\eq
Note that both $A^{exc}_{SHS}$ and $A^{exc}_{HS}$ depend upon
only a finite number of moments $\xi_{\nu}$, and this is the crucial
feature for the feasibility of the phase equilibrium, as remarked.

For the chemical potential $\beta \mu_i=\partial(\beta
A/V)/\partial\rho_i$ we find after some algebra
\bq \nonumber
\beta\mu^{exc}(\sigma,\tau,\rho;\{\xi_i\})&=&\left(\mu^{[0]}_{HS}+\Delta
\mu^{[0]}\right)+\left(\mu^{[1]}_{HS}+\Delta\mu^{[1]}\right)\sigma+\\
&&\left(\mu^{[2]}_{HS}+\Delta\mu^{[2]}\right)\sigma^2+
\left(\mu^{[3]}_{HS}+\Delta\mu^{[3]}\right)\sigma^3~,
\eq
where
\bq
\mu^{[0]}_{HS}&=&-\ln(1-\xi_3)~,\\
\mu^{[1]}_{HS}&=&3\xi_2/(1-\xi_3)~,\\
\mu^{[2]}_{HS}&=&\left(3\frac{\xi_2^2}{\xi_3^2}\right)\ln(1-\xi_3)+
3\xi_1/(1-\xi_3)+\left(3\frac{\xi_2^2}{\xi_3}\right)/(1-\xi_3)^2~,\\
\nonumber
\mu^{[3]}_{HS}&=&\left(-2\frac{\xi_2^3}{\xi_3^3}\right)\ln(1-\xi_3)+
\left(\xi_0-\frac{\xi_2^3}{\xi_3^2}\right)/(1-\xi_3)+
\left(3\xi_1\xi_2-\frac{\xi_2^3}{\xi_3^2}\right)/(1-\xi_3)^2+\\
&&\left(2\frac{\xi_2^3}{\xi_3}\right)/(1-\xi_3)^3~.
\eq
and
\bq
\Delta\mu^{[0]}=\left\{
\begin{array}{ll}
\displaystyle
\frac{1}{\tau}\frac{\xi_1^3}{\xi_0^2}&\mbox{Case I}~,\\
\displaystyle
0&\mbox{Case II, III}~,\\
\displaystyle
-\frac{1}{\tau}\frac{\xi_3}{4}&\mbox{Case IV}~,\\
\displaystyle
\frac{1}{\tau}\frac{\xi_1^3}{2\xi_0^2}&\mbox{Case V}~,
\end{array}\right.
\eq
\bq
\Delta\mu^{[1]}=\left\{
\begin{array}{ll}
\displaystyle
-\frac{1}{\tau}\frac{3\xi_1^2}{\xi_0}&\mbox{Case I}~,\\
\displaystyle
-\frac{1}{\tau}\xi_2&\mbox{Case II, III}~,\\
\displaystyle
-\frac{1}{\tau}\frac{3\xi_2}{4}&\mbox{Case IV}~,\\
\displaystyle
-\frac{1}{\tau}\frac{1}{2}\left(\xi_2+\frac{3\xi_1^2}{\xi_0}\right)
&\mbox{Case V}~,
\end{array}\right.
\eq
\bq
\Delta\mu^{[2]}=\left\{
\begin{array}{ll}
\displaystyle
0&\mbox{Case I}~,\\
\displaystyle
-\frac{1}{\tau}\xi_1&\mbox{Case II, III}~,\\
\displaystyle
-\frac{1}{\tau}\frac{3\xi_1}{4}&\mbox{Case IV}~,\\
\displaystyle
-\frac{1}{\tau}\frac{\xi_1}{2}&\mbox{Case V}~,
\end{array}\right.
\eq
\bq
\Delta\mu^{[3]}=\left\{
\begin{array}{ll}
\displaystyle
0&\mbox{Case I}~,\\
0&\mbox{Case II, III}~,\\
\displaystyle
-\frac{1}{\tau}\frac{\xi_0}{4}&\mbox{Case IV}~,\\
\displaystyle
0 &\mbox{Case V}~,
\end{array}\right.
\eq

It is noteworthy that if we retain in the expression
(\ref{polycoex:PHS}) for $P_{HS}$, only the first term, then our Case
IV coincides with the van der Waals model of Bellier-Castella 
{\sl et. al.} \cite{Bellier00} with $n=1, l=0$, upon identifying
$4\tau$ with the temperature used by these authors.

\subsubsection{C1 approximation with Case V}
\label{sec:C1binodal}
In analogy with what we have done before, we now consider the C1
approximation for Case V. Using Eq. (\ref{c1:y(sigma)}) into
Eq. (\ref{dadtau})  
\bq
\frac{\partial (\beta A^{exc}/N)}{\partial\tau}=12\frac{\eta}{\tau}\left[
k_0\frac{\langle\sigma\rangle}{\langle\sigma^3\rangle}+
k_1\left(\frac{\langle\sigma^2\rangle+\langle\sigma\rangle^2}
{\langle\sigma^3\rangle}\right)+
k_2\frac{\langle\sigma^2\rangle\langle\sigma\rangle}{\langle\sigma^3\rangle}
\right]~.
\eq
Integrating from $\tau=\infty$ we find
\bq \nonumber
\frac{\beta(A^{exc}_{SHS}-A^{exc}_{HS})}{N}&=&-\frac{\eta}{2}\frac{1}{\tau}
\left(\frac{\langle\sigma\rangle^3}{\langle\sigma^3\rangle}+
\frac{\langle\sigma\rangle\langle\sigma^2\rangle}{\langle\sigma^3\rangle}
\right)+\\ \nonumber
&&\frac{\eta^2}{2}\left[-\frac{1}{\tau}\left(
\frac{\langle\sigma\rangle\langle\sigma^2\rangle}{\langle\sigma^3\rangle}
+\frac{\langle\sigma\rangle^3}{\langle\sigma^3\rangle}
+3\frac{\langle\sigma\rangle^2\langle\sigma^2\rangle^2}{\langle\sigma^3\rangle^2}\right)\right.+\\
\nonumber
&&\frac{1}{\tau^2}\left(\frac{1}{4}
\frac{\langle\sigma\rangle^2\langle\sigma^2\rangle^2}{\langle\sigma^3\rangle^2}
+\frac{3}{4}
\frac{\langle\sigma\rangle^4\langle\sigma^2\rangle}{\langle\sigma^3\rangle^2}\right)-\\ \label{C1:bA/N}
&&\left.\frac{1}{\tau^3}\left(\frac{1}{72}
\frac{\langle\sigma\rangle^6}{\langle\sigma^3\rangle^2}
+\frac{1}{24}
\frac{\langle\sigma\rangle^4\langle\sigma^2\rangle}{\langle\sigma^3\rangle^2}\right)\right]~,
\eq

For this case we limit ourselves to study the coexistence problem
for the one-component system.
In table \ref{tab:crit} we compare the critical parameters obtained
through the energy route for the mMSA, C1, PY
approximations and MC simulation, for the one-component
system. 

Notice that, as already pointed out in Ref. \cite{Gazzillo04}, a density
expansion of $y(\sigma)$ within the PY approximation gives to zero-order
the $y(\sigma)$ of the mMSA approximation and to first-order the
$y(\sigma)$ of the C1 approximation (as should be expected comparing
the density expansions of the closures corresponding to these
approximations). So 
at low densities $Z_{SHS}$ from mMSA, C1, and PY should be
comparable. From table \ref{tab:crit} we see that the true critical
parameters are between the PY and the C1 ones.

In Fig. \ref{fig:binodal-c1} we depict the binodal curve obtained from
the C1 approximation for the one-component system and we compare it
with the PY binodal curve (obtained from the
energy route) \cite{Watts71} and the one resulting from the MC
simulation of Miller and Frenkel \cite{Miller04}. Remarkably, the gas-liquid
coexistence curve predicted by C1 lies closer to the MC data than the
one predicted by PY on the gas branch and further on the liquid branch.

\subsection{Numerical results}

In this section we describe the numerical results obtained from the
solution of Eqs. (\ref{tp:eq1})-(\ref{tp:eq3}) for the SHS in the
mMSA, through a Newton-Raphson algorithm.

We first determined the {\sl cloud} and {\sl shadow} curves by solving
Eqs. (\ref{tp:eq1})-(\ref{tp:eq3}) for the particular case in which we
set $\rho^{(0)}=\rho^{(1)}$ so that
$F^{(1)}(\sigma)=F^{(0)}(\sigma)$. The cloud curve $\rho_c(\tau)$ is   
such that the solutions $\rho^{(1)}(\tau)$, $\rho^{(2)}(\tau)$ of the
full coexistence problem given by Eqs. (\ref{tp:eq1})-(\ref{tp:eq3}),
for a fixed $\rho^{(0)}$ (the coexistence or binodal curves), have the
property that for a certain 
temperature $\tau_0$, $\rho^{(1)}(\tau_0)=\rho_c(\tau_0)=\rho^{(0)}$,
i.e the density of phase 1 ends on the cloud curve. The shadow curve
is the set of points $\rho_s(\tau)$ in equilibrium with the
corresponding cloud curve, i.e. $\rho^{(2)}(\tau_0)=\rho_s(\tau_0)$,
the density of phase 2 ends on the shadow curve. The interception
between the cloud and the corresponding shadow curve gives the critical
point $(\tau_{cr},\rho_{cr})$: when $\rho^{(0)}=\rho_{cr}$ the two
solutions $\rho^{(1)}(\tau)$, $\rho^{(2)}(\tau)$ meet at the critical
point.  

In order to find the cloud and shadow curves we choose as the parent
distribution $F^{(0)}(\sigma)$ a Schulz distribution with
$\langle\sigma\rangle$=1, and the initial conditions, for the
Newton-Raphson algorithm, in the high
temperature $\tau_*$ and low polydispersity $s_*$ region. Our
starting conditions for the solution are
\bq
\rho^{(\alpha)}&=&\rho^{(\alpha)}_{oc}~,\\
\xi_1^{(\alpha)}&=&\frac{\pi}{6}\rho^{(\alpha)}~,\\
\xi_2^{(\alpha)}&=&\frac{\pi}{6}\rho^{(\alpha)}(1+s^2_*)~,\\
\xi_3^{(\alpha)}&=&\frac{\pi}{6}\rho^{(\alpha)}(1+s^2_*)(1+2s^2_*)~,
\eq 
for $\alpha=1,2$, where $\rho^{(1)}_{oc}$ and $\rho^{(2)}_{oc}$ are
the coexistence densities at a temperature $\tau_*$ for the one
component system. Once the cloud and shadow curves are determined we
proceed to find the coexistence curves for a given mother density.

In Fig. \ref{fig:pd-m1} we depict the cloud and shadow
curves obtained with our Case I for two representative
values of polydispersity, $s=0.1$ and $s=0.3$. For
comparison the coexistence curve of the one component
system ($s=0$) is also reported. As polydispersity increases, the
critical point moves 
to lower densities and lower temperatures ($\tau_{cr}\simeq
0.094,\rho_{cr}\simeq 0.249$ at $s=0$, $\tau_{cr}\simeq
0.093,\rho_{cr}\simeq 0.24$ at $s=0.1$, and $\tau_{cr}\simeq
0.085,\rho_{cr}\simeq 0.197$ at $s=0.3$). 
Let us now fix $s=0.3$, a value corresponding to a moderate polydispersity.
Again in Fig. \ref{fig:pd-m1} we depict 
three coexistence curves upon changing the mother density  $\rho^{(0)}=0.08$,
$\rho^{(0)}=0.25$, and $\rho^{(0)}=0.197\simeq\rho_{cr}$.

All these curves closely resemble the corresponding ones obtained for the
polydisperse van der Waals model \cite{Bellier00}, in agreement with
previous results.
In Fig. \ref{fig:dist-m1} we show how the two daughter distribution
functions (at $s=0.3$ and  $\rho^{(0)}=\rho_{cr}$) differ from the
parent Schulz distribution (a process usually called {\sl
fractionation}), for two different values of
temperature $\tau=0.084$ and $\tau=0.078$.

Next we consider differences in the critical behavior with respect to
changement in the Case.
In Fig. \ref{fig:cs-s.3} we show the cloud and shadow curves obtained
using Case I, IV, and V at $s=0.3$. While for Case I and V the
critical point is displaced at lower temperature and lower density
respect to the monodisperse system, the critical point of
Case IV is displaced at higher temperatures and lower density. The
cloud curves of Case II and III have a low density branch that does not
meet the high density one as soon as $s>0$; moreover, the cloud curve
does not meet the corresponding shadow curve, so there is no critical point. 
We are not aware of similar features in other polydisperse models, although
this is of course to be expected in other cases as well.

\section{Conclusions}
\label{sec:concl}

In this work we have performed an extensive analysis of the phase
diagram for Baxter SHS 
model in the presence of polydispersity. In spite of its simplicity, this
model has been proven to be extremely useful in the theoretical
characterization of sterically stabilized colloids. These systems are,
however, affected by intrinsic polydispersity in some of their physical 
properties (size, species, etc) and hence the effect of polydispersity
on the corresponding theoretical
models cannot be overlooked and is then a rather interesting point to address.
As only formal manipulations \cite{Barboy79} can be carried out for the
multicomponent 
Baxter SHS model within the PY approximation, we have resorted to
a simpler closure (mMSA) to which the PY closure reduces in the limit of
zero density and that was recently shown \cite{Gazzillo04} to
reproduce rather precisely many of the interesting features of its PY
counterpart. 
Our analysis has also been inspired
by recent results by Miller and Frankel \cite{Miller04} who showed that
Baxter SHS model coupled with PY closure reproduced fairly well
their MC data in the one-component case.
We have studied the effect of polydispersity on 
phase stability boundaries, the
percolation phase transition, and the gas-liquid phase transition.
We have considered five different cases of polydispersity. This is because
there is no general agreement on the way in which adhesion forces
are depending on the size of particles.
Case I and II had already been discussed in previous work by us
\cite{Gazzillo00}, Case III is a variant of Case I, whereas a case
similar to Case IV had been employed by Tutschka and Kahl
\cite{Tutschka98}.  
Finally Case V has been specifically devised to cope with approximation C1.
In spite of the apparent redundancy of all these sub-cases, we believe that
each of these examples has a reasonable interest on its own, and hence 
we have included them all in our discussion. 

We studied the instability boundaries and the two-phase coexistence
problem of polydisperse SHS system 
in the mMSA (C0). The next level of approximation (C1) would still be feasible,
but significantly more lengthly. We have laid down the necessary formalism in
Sections \ref{sec:c1} and \ref{sec:C1binodal}, and tested its effect on the 
one-component case, by comparing the results against the PY
approximation and MC data.
We derived the percolation threshold of the polydisperse system both within 
mMSA (C0) closure (for all five Cases) and in the C1 approximation
(using Case V). 

We found that the effect of polydispersity on the stability and phase
boundaries slightly depends upon the chosen Case, but there are
general features shared by all of them:
polydispersity renders the mixture more stable
with respect to concentration fluctuations (in the small density
region, see Fig. \ref{fig:mech}) with the exception of Case II for which
the stability boundary is independent from the polydispersity; 
Eqs. (\ref{mMSA:percolation}), (\ref{mMSA:perc-modelIV}), and
(\ref{mMSA:perc-modelV}) (in the mMSA), and Eq. (\ref{C1:perctrans})
(in the C1), describe its effect on the percolation threshold (see
Figs. \ref{fig:binodal-c1} and \ref{fig:percolation-c1-poly}). 
Polydispersity increases the region of the phase diagram where we have
a non-percolating phase, with the exception of Case
IV, for which the opposite trend is observed, and of Case II for which
the percolation threshold is independent from the polydispersity;
polydispersity reduces the region of the phase diagram where we
have a gas-liquid coexistence for Cases I and V, while the opposite
trend is observed for Case IV (see Fig. \ref{fig:cs-s.3}). For Case
II and III we obtained cloud curves with a gap at high temperature,
moreover the cloud curve does not meet the corresponding shadow curve,
so there is no critical point, as soon as polydispersity is
introduced. 

In conclusion, the typical effect of polydispersity is to reduce the
size of the unstable region, the percolating region, and the two-phase
region of the phase diagram, although exceptions to this
general rule have been observed for Case II, III, and IV. 

For the one-component case we also compared the percolation threshold
and binodal curve obtained from the C1 approximation with the results 
from the PY approximation \cite{Chiew83,Watts71} and the results
from the Monte Carlo simulation of Miller and Frenkel \cite{Miller04}
(see Fig. \ref{fig:binodal-c1}). The percolation threshold from C1
appears to approach that from PY, but is still significantly different
from the results from the Monte Carlo simulation (the zero density
limit, on the other hand, appears to be more physically sound than the
PY one, and this feature remains to be elucidated). 
The gas-liquid coexistence curve predicted by C1 
is better than the one given by PY on the gas branch and worse on
the liquid branch. Table \ref{tab:crit} shows how the true (from the
Monte Carlo simulation of Miller and Frenkel \cite{Miller04}) critical
temperature and density for the gas-liquid coexistence should lay
between the ones predicted by PY and the ones predicted by C1.

Future developments of the present work can be envisaged along the
following lines: (i) as pointed out in \cite{Gazzillo94} on
defining $\psi_G=\prod_m x_m/S_{{\rm CC}}(0)$ and 
$\psi_{\hat{A}}=\prod_m x_m/[(\rho k_B TK_T)S_{{\rm CC}}(0)]$, the
condition $\psi_G>0$ is necessary but not sufficient for the material
stability of the system and the condition $\psi_{\hat{A}}>0$ is
necessary but not sufficient for the mixed material and mechanical
stability. It could 
happen that those two conditions are satisfied but the system is
nonetheless unstable as occur for example in the binary mixture
studied by Chen and Forstmann \cite{Chen92}. Even though a
characterization of the instability boundary in the spirit of Chen and
Forstmann seems unattainable for a polydisperse system, it would be
desirable, in the future, a more precise location of the
instability boundaries. Moreover the way we found the instability
boundaries for the polydisperse system was to start from the
instability condition valid for a discrete mixture and take the limit
of a continuous mixture on the instability boundaries of the discrete
mixture. It would be interesting to compare our analysis with the one
given by Bellier-Castella {\sl et. al} (see section II C in
Ref. \cite{Bellier00}) who take the continuous limit from the outset; 
(ii) all the percolation thresholds that we
have calculated have a low density branch that enters the gas-liquid
coexistence region. The same feature is observed for the one-component
system studied through Monte Carlo simulation
\cite{Miller03,Miller04}. While it is clear that continuum percolation
is, strictly speaking, not a thermodynamic phase transition, 
one could expect, from a ``dynamical'' point of view, an interference
between the formation of infinite clusters of particles and phase
separation, and a clarification of this point would have interesting
experimental applications; (iii) the polydisperse system is expected
to display, in the low temperature region, other critical points
signaling the onset of $m>2$ phase coexistence \cite{Bellier00}, and
it would be interesting to study their evolution with polydispersity
for our system. 

\appendix

\section*{Appendix A: Determinant and inverse of a dyadic matrix}
\label{app:dyadic}
\renewcommand{\theequation}{A.\arabic{equation}}

Given the $n$-dyadic matrix of Eq. (\ref{s6}), its determinant is
\begin{equation}
\left| \widehat{{\bf Q}}\right| =\left| \ 
\begin{array}{cccc}
~1+{\bf A}^{(1)}\cdot {\bf {B}}^{(1)} & \qquad {\bf A}^{(1)}\cdot {\bf {B}}
^{(2)} & \cdots & \qquad {\bf A}^{(1)}\cdot {\bf {B}}^{(n)} \\ 
\qquad {\bf A}^{(2)}\cdot {\bf {B}}^{(1)} & \ 1+{\bf A}^{(2)}\cdot {\bf {B}}
^{(2)} & \cdots & \qquad {\bf A}^{(2)}\cdot {\bf {B}}^{(n)} \\ 
\vdots & \vdots & \vdots & \vdots \\ 
\qquad {\bf A}^{(n)}\cdot {\bf {B}}^{(1)} & \qquad {\bf A}^{(n)}\cdot {\bf {B%
}}^{(2)} & \cdots & 1+{\bf A}^{(n)}\cdot {\bf {B}}^{(n)}
\end{array}
\right|~.  \label{s7}
\end{equation}
Furthermore, any dyadic matrix $\widehat{{\bf Q}}$ admits {\it analytic}
inverse for {\it any } number $p$ of components, with elements given by
\begin{equation}
\widehat{Q}_{ij}^{-1}=\frac{1}{\left| \widehat{{\bf Q}}\right| }\left| 
\begin{array}{lllll}
\delta _{ij}\quad & \ \qquad B_{j}^{(1)} & \qquad B_{j}^{(2)} & \cdots & 
\qquad B_{j}^{(n)} \\ 
A_{i}^{(1)}\quad & ~1+{\bf A}^{(1)}\cdot {\bf {B}}^{(1)} & \qquad {\bf A}
^{(1)}\cdot {\bf {B}}^{(2)} & \cdots & \qquad {\bf A}^{(1)}\cdot {\bf {B}}
^{(n)} \\ 
A_{i}^{(2)} & \qquad {\bf A}^{(2)}\cdot {\bf {B}}^{(1)}{\bf \quad } & \ 1+
{\bf A}^{(2)}\cdot {\bf {B}}^{(2)} & \cdots & \qquad {\bf A}^{(2)}\cdot {\bf 
{B}}^{(n)} \\ 
\ \vdots & \ \qquad \quad \vdots & \ \qquad \quad \vdots & ~\vdots & \
\qquad \quad \vdots \\ 
A_{i}^{(n)} & \qquad {\bf A}^{(n)}\cdot {\bf {B}}^{(1)} & \qquad {\bf A}
^{(n)}\cdot {\bf {B}}^{(2)} & \cdots & 1+{\bf A}^{(n)}\cdot {\bf {B}}^{(n)}
\end{array}
\right|~.  \label{s7b}
\end{equation}

\section*{Appendix B: Derivation of Eq. (\ref{mMSA:q+})}
\label{app:q+}
\renewcommand{\theequation}{B.\arabic{equation}}

The closure condition (\ref{mMSA:c+closure}) justify the usual
generalized Wiener-Hopf factorization \cite{Baxter70}
\bq
rc^+_{ij}(|r|)&=&-{q^+_{ij}}^\prime(r)+2\pi\sum_m\rho_m\int_{L_{mi}}^\infty
dt q^+_{mi}(t){q^+_{mj}}^\prime(r+t)~,\\ \label{baxter:h+}
rh^+_{ij}(|r|)&=&-{q^+_{ij}}^\prime(r)+2\pi\sum_m\rho_m\int_{L_{im}}^\infty
dt q^+_{im}(t)(r-t)h^+_{mj}(|r-t|)~,\\
\eq
where $r>L_{ij}$, the prime denotes differentiation, and $q^+_{ij}(r)$
are real functions with support on $[L_{ij},\sigma_{ij}]$ and zero
everywhere else.

Let us determine $q^+_{ij}(r)$. Using the exact condition
(\ref{mMSA:h+closure}) in Eq. (\ref{baxter:h+}) we find for
$L_{ij}<r\le\sigma_{ij}$ 
\bq
{q^+_{ij}}^\prime(r)=-K_{ij}\delta(|r|-\sigma_{ij})+2\pi\sum_m\rho_m
\int_{L_{im}}^{\sigma_{im}}dt q^+_{im}(t)(r-t)\frac{K_{mj}}{\sigma_{mj}}
\delta(|r-t|-\sigma_{mj})~.
\eq
The second term on the right end side is equal to $2\pi\sum_m\rho_m
q^+_{im}(r-\sigma_{mj})K_{mj}$ which is zero when $r<\sigma_{ij}$. So
we simply have
\bq
{q^+_{ij}}^\prime(r)=-K_{ij}\delta(|r|-\sigma_{ij})~~~~L_{ij}<r\le\sigma_{ij}~.
\eq
Integrating this equation gives Eq. (\ref{mMSA:q+}).

\section*{Appendix C: Coefficients of Eq. (\ref{C1:perctrans})}
\label{app:c1-perc}
\renewcommand{\theequation}{C.\arabic{equation}}

The coefficients in Eq. (\ref{C1:perctrans}) are as follows 
\bq
q_1(s,\eta)&=&-\frac{\eta\,\left( 2 + 5\,\eta \right) \,{\left( 1 +
    3\,s^2 + 2\,s^4 \right) }^3}{2\,{\left( 1 + s^2 \right)
  }^3\,{\left( 1 + 2\,s^2 \right)}^4}~,\\
q_2(s,\eta)&=&-\frac{\eta^2\,\left\{- 4 + 
       \left[ \eta\,\left( 2 + \eta \right) - 5 \right] \,s^2 \right\} \,
     {\left( 1 + 3\,s^2 + 2\,s^4 \right) }^2}{4\,{\left( 1 + s^2 \right) }^3\,
     {\left( 1 + 2\,s^2 \right) }^4}~,\\
q_3(s,\eta)&=&\frac{\eta^2\,\left\{ -2+\left[ 6\,\eta\,\left( 1
  + \eta \right) -5 \right] \,s^2 - 2\,s^4 \right\} }{24\,\left( 1 + s^2
  \right) \,{\left( 1 + 2\,s^2 \right) }^3}~,\\
q_4(s,\eta)&=&-\frac{\eta^3\,s^2\,\left[ 2 + 5\,\eta + \left( 4 +
  7\,\eta \right) \,s^2 \right] }{96\,{\left( 1 + s^2 \right)
  }^2\,{\left( 1 + 2\,s^2 \right)}^4}~,\\
q_5(s,\eta)&=&0~,\\
q_6(s,\eta)&=&\frac{\eta^4\,s^4}{2304\,{\left( 1 + s^2 \right)
  }^3\,{\left( 1 + 2\,s^2 \right) }^4}~. 
\eq

\section*{Appendix D: Phase coexistence conditions}
\label{app:coex}
\renewcommand{\theequation}{D.\arabic{equation}}

In this Appendix we give a complete derivation of
Eqs. (\ref{tp:eq1})-(\ref{tp:eq3}) in the main text.

Consider a $p-$component mixture. Each species $i$ has number density
$\rho_i^{(0)}=N_i^{(0)}/V^{(0)}$, where $N_i^{(0)}$ is the number of
particles of type $i$ and $V^{(0)}$ the volume of the system.

We assume that at a certain temperature $\tau$ the system separates
into $m$ daughter phases, where each phase $\alpha=1,\ldots,m$ is
characterized by a volume $V^{(\alpha)}$ and a number of particles of
species $i$, $N_i^{(\alpha)}$.

At equilibrium the following set of constraints must be fulfilled:
\begin{itemize}
\item[(1)] volume conservation
\bq\label{equi1}
V^{(0)}=\sum_{\alpha=1}^m V^{(\alpha)}~,
\eq
\item[(2)] conservation of the total number of particles of each
species
\bq\label{equi2}
N_i^{(0)}=\sum_{\alpha=1}^m N_i^{(\alpha)}~,~~~i=1,\ldots,p~,
\eq
\item[(3)] equilibrium condition for the pressures
\bq\label{equi3}
P^{(\alpha)}(\tau,V^{(\alpha)},\{N_i^{(\alpha)}\})= 
P^{(\beta)}(\tau,V^{(\beta)},\{N_i^{(\beta)}\})~,
\eq
\item[(4)] equilibrium condition for the chemical potentials
\bq\label{equi4}
\mu_i^{(\alpha)}(\tau,V^{(\alpha)},\{N_i^{(\alpha)}\})=
\mu_i^{(\beta)}(\tau,V^{(\beta)},\{N_i^{(\beta)}\})~,~~~i=1,\ldots,p~,
\eq
\end{itemize}
where $\{N_i^{\alpha}\}$ is a short-hand notation for 
$N_1^{\alpha},\ldots,N_p^{\alpha}$.

It is convenient to use the following set of variables: $\tau$; 
$\rho^{(\alpha)}=N^{(\alpha)}/V^{(\alpha)}$;  
$x_i^{(\alpha)}=N_i^{(\alpha)}/N^{(\alpha)}$, $i=1,\ldots,p$ with 
$N^{(\alpha)}=\sum_i N_i^{(\alpha)}$. Introducing
$x^{(\alpha)}=N^{(\alpha)}/N^{(0)}$ Eqs. (\ref{equi1})-(\ref{equi4})
can be rewritten as follows
\bq\label{equi1b}
\frac{1}{\rho^{(0)}}&=&\sum_\alpha
\frac{1}{\rho^{(\alpha)}}x^{(\alpha)}~,\\\label{equi2b}
x_i^{(0)}&=&\sum_\alpha x_i^{(\alpha)}x^{(\alpha)}~,\\\label{equi3b}
P^{(\alpha)}(\tau,\rho^{(\alpha)},\{x_i^{(\alpha)}\})&=&
P^{(\beta)}(\tau,\rho^{(\beta)},\{x_i^{(\beta)}\})~,\\\label{equi4b}
\mu_i^{(\alpha)}(\tau,\rho^{(\alpha)},\{x_i^{(\alpha)}\})&=&
\mu_i^{(\beta)}(\tau,\rho^{(\beta)},\{x_i^{(\beta)}\})~,
\eq
with the normalization condition
\bq
\label{normb}
\sum_ix_i^{(\alpha)}=1~,~~~\alpha=1,\ldots,m~.
\eq
Eqs. (\ref{equi1b})-(\ref{normb}) form a set of closed equations for
the $(2+p)m$ unknowns $\rho^{(\alpha)}$, $x^{(\alpha)}$,
$x_i^{(\alpha)}$ with $i=1,\ldots,p$ and $\alpha=1,\ldots,m$.
Notice that when $m=p+1$ the densities of the various phases
$\rho^{(\alpha)}$ will be independent of $\rho^{(0)}$, since
relations (\ref{equi3b}), (\ref{equi4b}), and (\ref{normb}) form a closed
set of equations for the unknowns $\rho^{(\alpha)}$, $x_i^{(\alpha)}$.

In the continuous polydisperse limit ($p\to\infty$) we have to take
into account 
the substitution rule (\ref{polycoex:dist}). Then the thermodynamic
quantities will be rewritten as in Eqs. (\ref{polycoex:P}) and
(\ref{polycoex:mu}), and Eqs. (\ref{equi1b})-(\ref{equi4b}) become
\bq\label{equi1c}
\frac{1}{\rho^{(0)}}&=&\sum_\alpha
\frac{1}{\rho^{(\alpha)}}x^{(\alpha)}~,\\\label{equi2c}
F^{(0)}(\sigma)&=&\sum_\alpha F^{(\alpha)}(\sigma)x^{(\alpha)}~,\\
\label{equi3c}
P^{(\alpha)}(\tau,\rho^{(\alpha)};[F^{(\alpha)}])&=&
P^{(\beta)}(\tau,\rho^{(\beta)};[F^{(\beta)}])~,\\\label{equi4c}
\mu^{(\alpha)}(\sigma,\tau,\rho^{(\alpha)};[F^{(\alpha)}])&=&
\mu^{(\beta)}(\sigma,\tau,\rho^{(\beta)};[F^{(\beta)}])~,
\eq
with the normalization condition
\bq \label{normc}
\int F^{(\alpha)}(\sigma)d\sigma=1~,~~~\alpha=1,\ldots,m~.
\eq
Integrating Eq. (\ref{equi2c}) over $\sigma$ and using
Eq. (\ref{normc}) we find 
\bq \label{equi5c}
\sum_\alpha x^{(\alpha)}=1~.
\eq 
The set of Eqs. (\ref{equi1c})-(\ref{normc}) form a closed set of
equations for the unknowns $\rho^{(\alpha)}$, $x^{(\alpha)}$, and
$F^{(\alpha)}(\sigma)$ with $\alpha=1,\ldots,m$. Notice that, due to
the substitution rule (\ref{polycoex:dist}), sum over $i$ becomes
integration over $\sigma$ and thermodynamic quantities become
functionals of the distribution function. We have indicated such
dependence with square brackets.

\subsection*{Two-phase coexistence}
\label{subsec:tp}

Let us now specialize ourselves to the two-phase ($m=2$)
coexistence. We are thus concentrating on the high temperature portion
of the phase diagram, since coexistence with $m>2$ (Gibbs phase rule
does not restrict the value of $m$ in a system of infinitely many
species) is expected to occur at low temperatures. From
Eqs. (\ref{equi5c}) and (\ref{equi1c}) we find 
\bq \label{tp:conc}
x^{(1)}&=&\frac{\rho^{(0)}-\rho^{(2)}}{\rho^{(1)}-\rho^{(2)}}
\frac{\rho^{(1)}}{\rho^{(0)}}~,\\
x^{(2)}&=&\frac{\rho^{(1)}-\rho^{(0)}}{\rho^{(1)}-\rho^{(2)}}
\frac{\rho^{(2)}}{\rho^{(0)}}~.
\eq
Notice that $x^{(1)}$ and $x^{(2)}$ must be positive. So if
$\rho^{(1)}<\rho^{(2)}$, then $\rho^{(0)}$ must lie between
$\rho^{(1)}$ and $\rho^{(2)}$, if $\rho^{(2)}<\rho^{(1)}$, it must lie
between $\rho^{(2)}$ and $\rho^{(1)}$.
Substituting these expressions in Eq. (\ref{equi2c}) we find
\bq \label{tp:F1}
\rho^{(2)}F^{(2)}=\rho^{(0)}F^{(0)}
\frac{\rho^{(1)}-\rho^{(2)}}{\rho^{(1)}-\rho^{(0)}}+
\rho^{(1)}F^{(1)}
\frac{\rho^{(0)}-\rho^{(2)}}{\rho^{(0)}-\rho^{(1)}}~.
\eq

Next we divide the chemical potentials in their ideal and excess
components $\mu=\mu^{id}+\mu^{exc}$ where
\bq
\beta{\mu^{id}}^{(\alpha)}(\sigma,\tau,\rho^{(\alpha)};[F^{(\alpha)}])=
\ln[\Lambda^3\rho^{(\alpha)}F^{(\alpha)}(\sigma)]~,
\eq
with $\Lambda$ being the de Broglie thermal wavelength. Now
Eq. (\ref{equi4c}) becomes
\bq \label{tp:F2}
F^{(1)}(\sigma)&=&F^{(2)}(\sigma)\frac{\rho^{(2)}}{\rho^{(1)}}
e^{\beta\Delta\mu^{exc}}~,\\
\Delta\mu^{exc}&=&{\mu^{exc}}^{(2)}(\sigma,\tau,\rho^{(2)};[F^{(2)}])-
{\mu^{exc}}^{(1)}(\sigma,\tau,\rho^{(1)};[F^{(1)}])~.
\eq

From Eqs. (\ref{tp:F1}) and (\ref{tp:F2}) we find
\bq \label{tp:F3}
F^{(\alpha)}(\sigma)=F^{(0)}(\sigma)Q^{(\alpha)}(\sigma,\tau,\rho^{(0)},
\rho^{(1)},\rho^{(2)};[F^{(1)}],[F^{(2)}])~,
\eq
where the $Q^{(\alpha)}$ are defined by Eq. (\ref{tp:Q}).

Formally the set of Eqs. (\ref{tp:F1}), (\ref{tp:F2}),
(\ref{equi3c}) with $\alpha=1,\beta=2$, and (\ref{normc}) with
$\alpha=1$ or 2, form a closed set of equations for the unknowns
$\rho^{(1)}, \rho^{(2)}, F^{(1)}(\sigma)$ and $F^{(2)}(\sigma)$.
In our case the free energy of the system [Case I, II, III, IV, and V
treated with mMSA, see Eq. (\ref{mMSA:bA/N}), or Case V treated with
C1, see Eq. (\ref{C1:bA/N})] is {\sl truncatable}: it is  
only a function of the three moments $\xi_i, i=1,2,3$ of the size
distribution function $F$. So the problem is mapped in the solution of 
the 8 Eqs. (\ref{tp:eq1})-(\ref{tp:eq3}) in the 8 unknowns
$\rho^{(1)}, \rho^{(2)}, \xi_1^{(1)}, 
\xi_2^{(1)}, \xi_3^{(1)}, \xi_1^{(2)}, \xi_2^{(2)}, \xi_3^{(2)}$.

\begin{figure}[hbtp]
\begin{center}
\includegraphics[width=16cm]{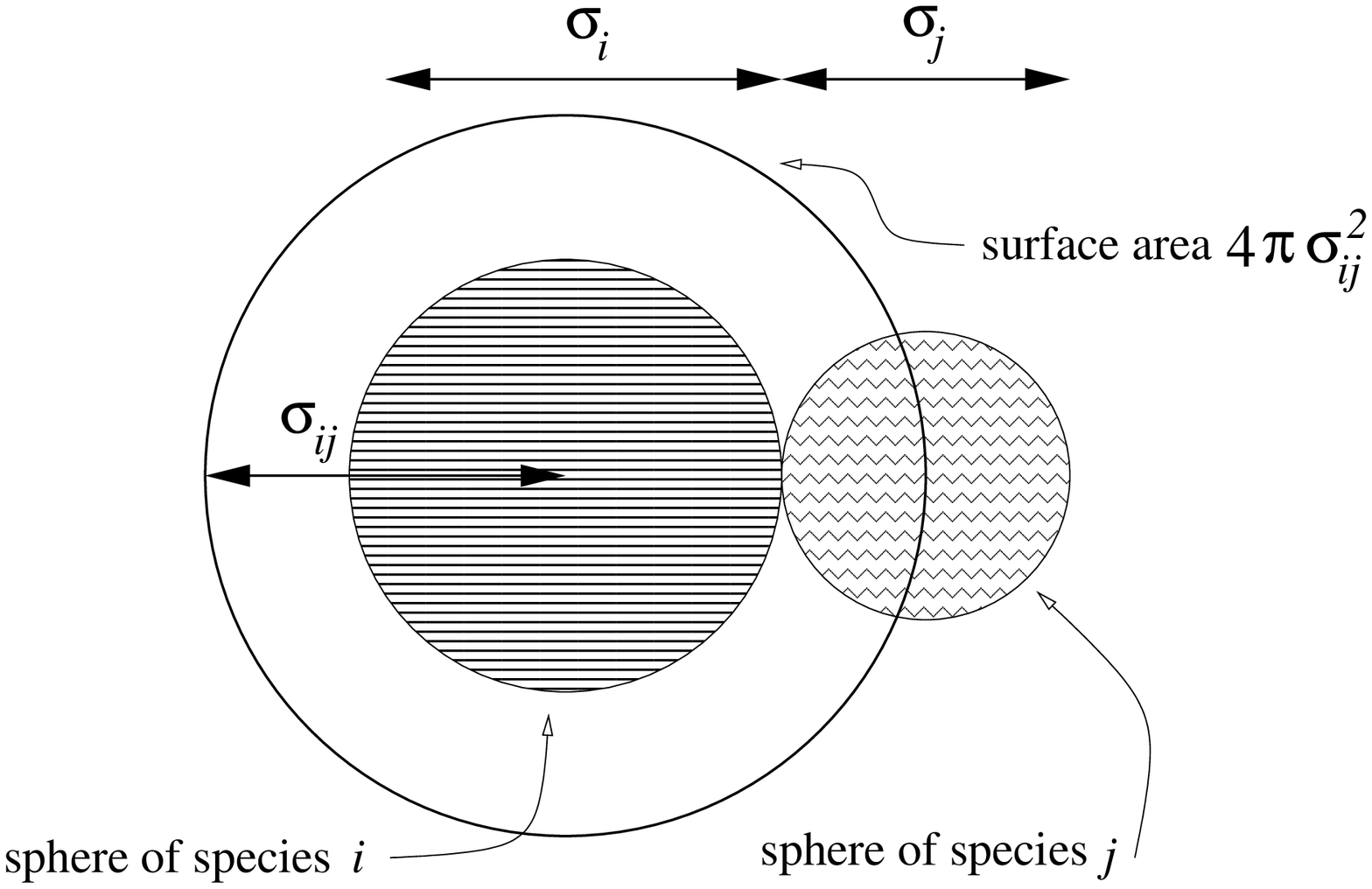}
\end{center}
\caption[]{R. Fantoni, D. Gazzillo, and A. Giacometti
\label{fig:p-cases}
}
\end{figure}
\begin{figure}[hbtp]
\begin{center}
\includegraphics[width=16cm]{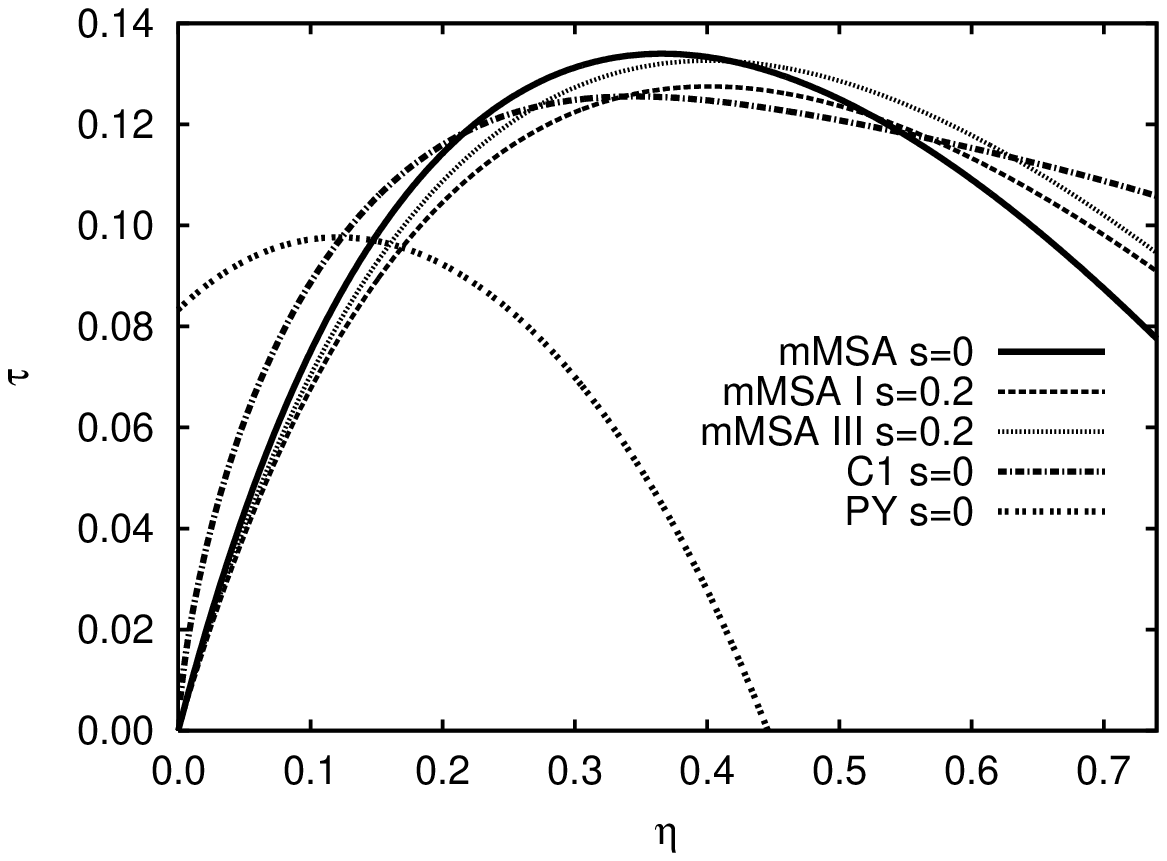}
\end{center}
\caption[]{R. Fantoni, D. Gazzillo, and A. Giacometti
\label{fig:mech}
}
\end{figure}
\begin{figure}[hbtp]
\begin{center}
\includegraphics[width=16cm]{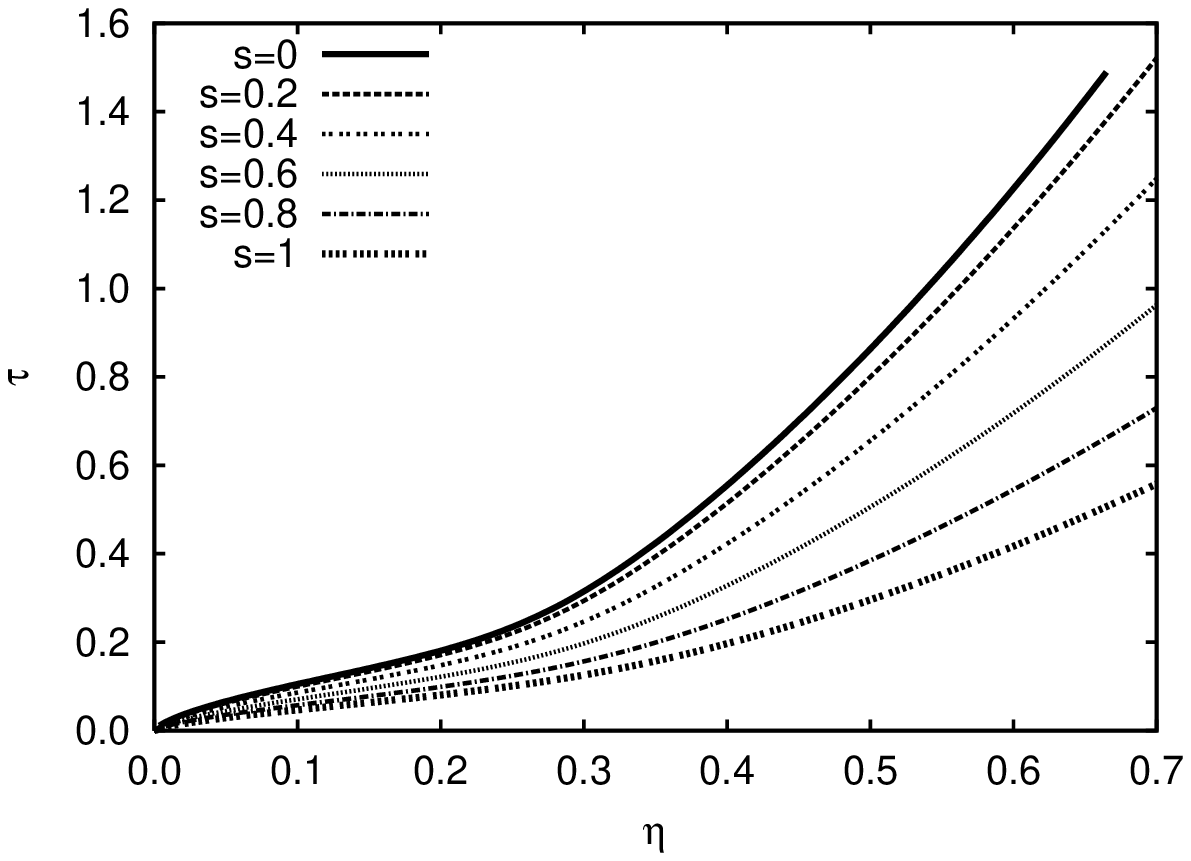}
\end{center}
\caption[]{R. Fantoni, D. Gazzillo, and A. Giacometti
\label{fig:percolation-c1-poly}
}
\end{figure}
\begin{figure}[hbtp]
\begin{center}
\includegraphics[width=16cm]{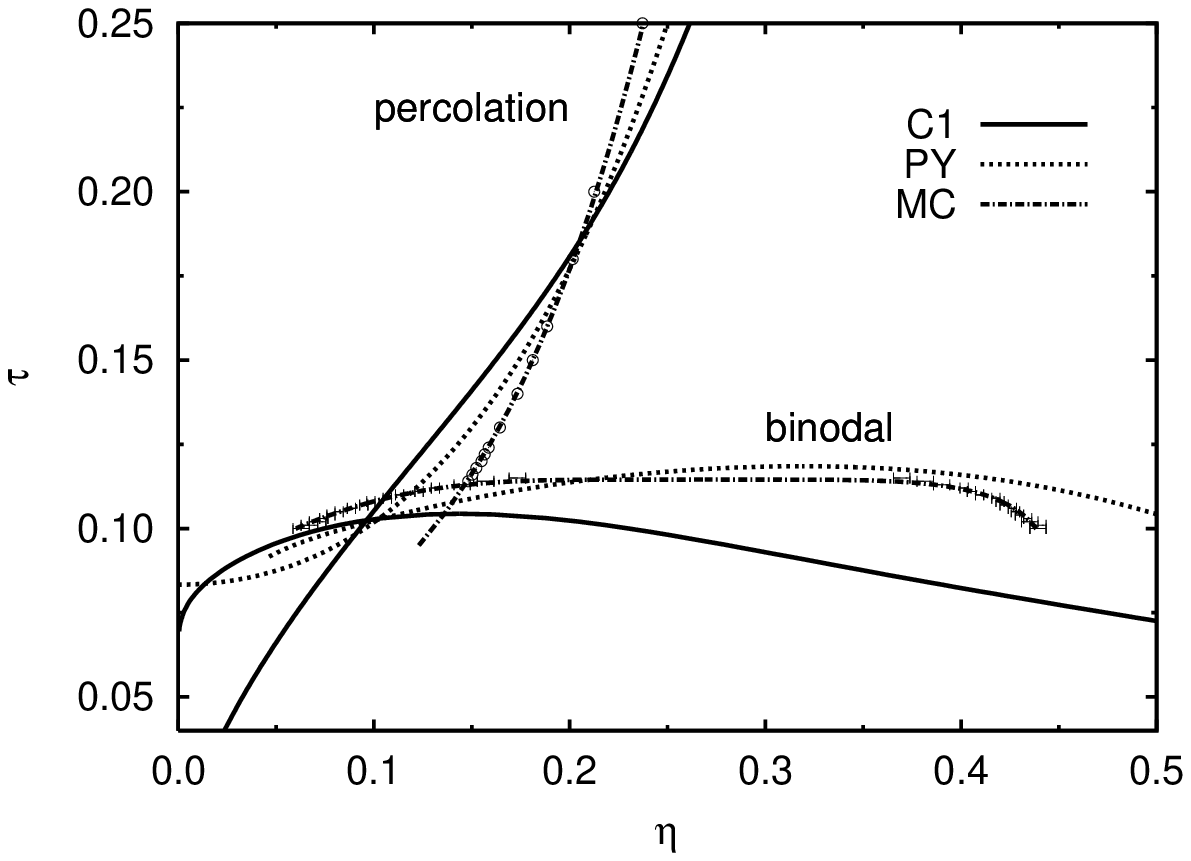}
\end{center}
\caption[]{R. Fantoni, D. Gazzillo, and A. Giacometti
\label{fig:binodal-c1}
}
\end{figure}
\begin{figure}[h!]
\begin{center}
\includegraphics[width=16cm]{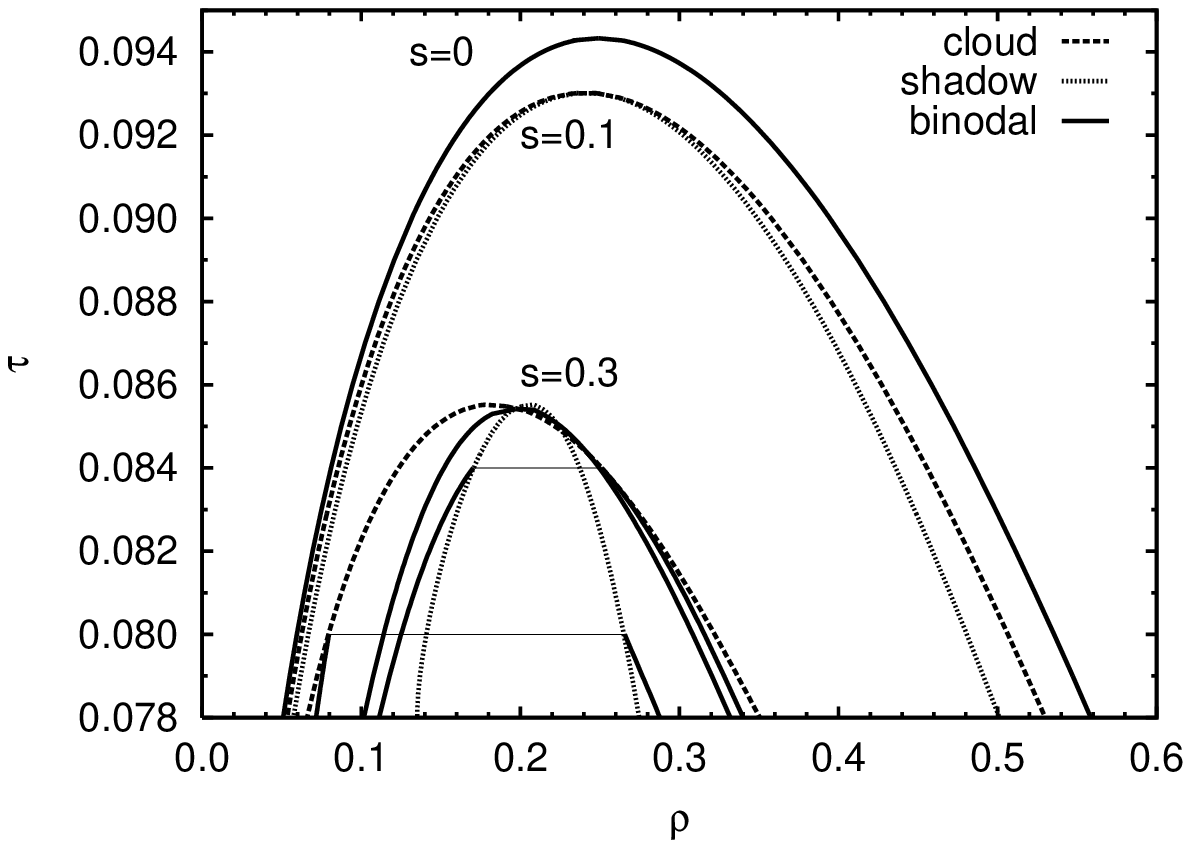}
\end{center}
\caption[]{R. Fantoni, D. Gazzillo, and A. Giacometti
\label{fig:pd-m1}
}
\end{figure}
\begin{figure}[h!]
\begin{center}
\includegraphics[width=16cm]{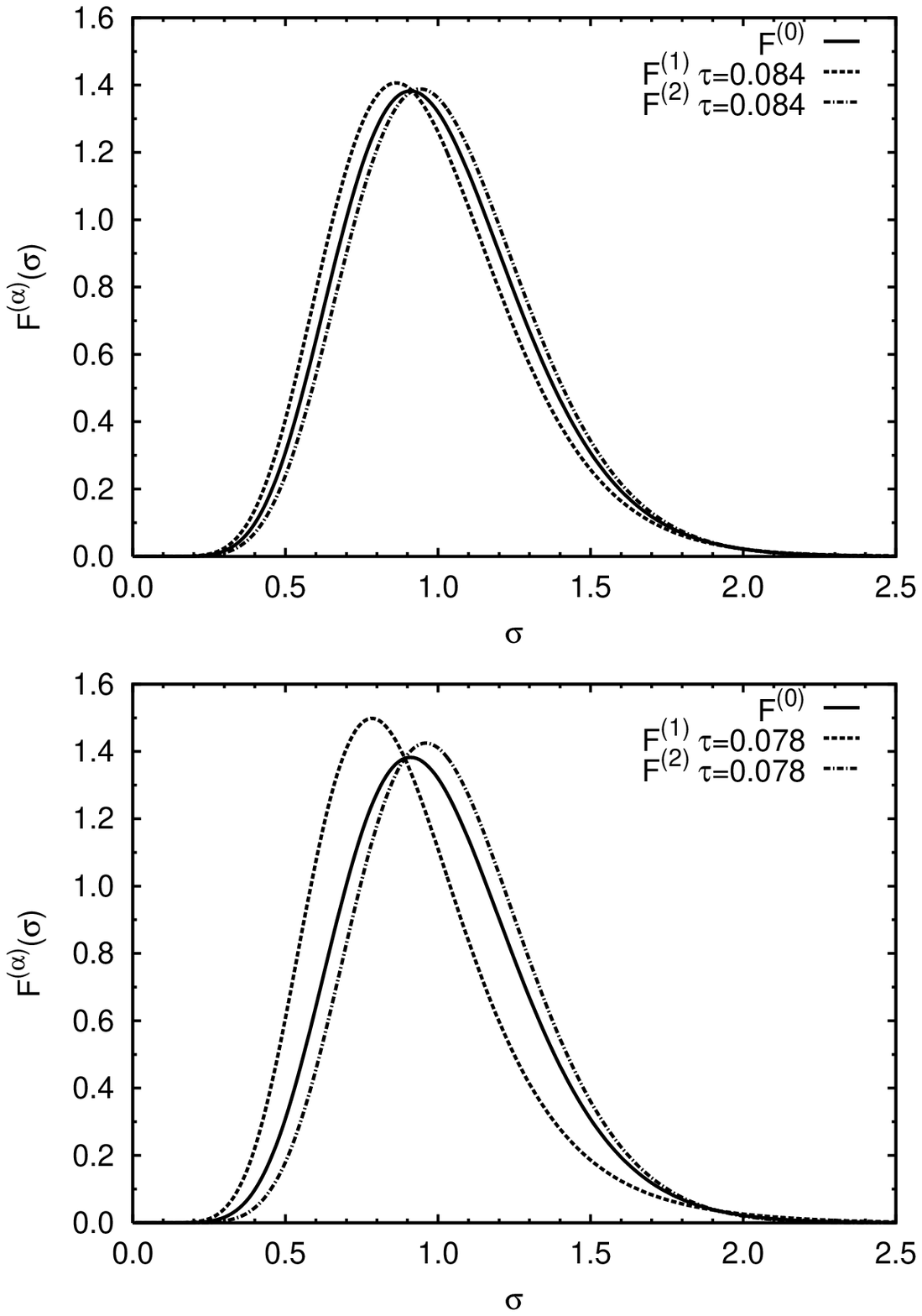}
\end{center}
\caption[]{R. Fantoni, D. Gazzillo, and A. Giacometti
\label{fig:dist-m1}
}
\end{figure}
\begin{figure}[h!]
\begin{center}
\includegraphics[width=16cm]{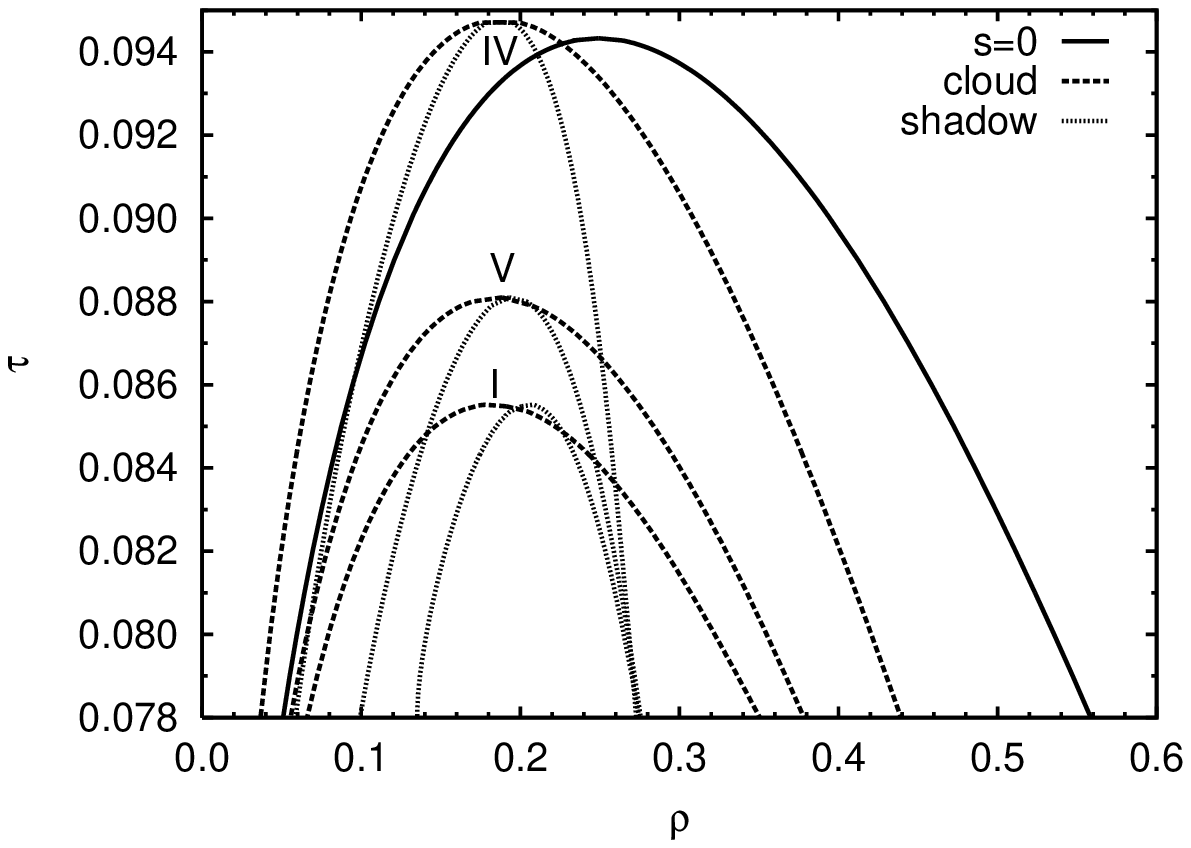}
\end{center}
\caption[]{R. Fantoni, D. Gazzillo, and A. Giacometti
\label{fig:cs-s.3}
}
\end{figure}

\bibliography{514502JCP}
\bibliographystyle{unsrt}
\newpage
\centerline{\bf LIST OF FIGURES}
\begin{itemize}

\item[Fig. \ref{fig:p-cases}] Schematic diagram showing the area of
the contact surface between a particle of species $i$ and a particle
of species $j$.

\item[Fig. \ref{fig:mech}] Curves for the onset of
phase instability (the fluid is stable above the
curves shown) as obtained from the mMSA approximation for a
monodisperse $s=0$ system, and for a polydisperse system with
$s=0.2$, and polydispersity chosen as in Cases I, II, and III [see
Eq. (\ref{s11})]. We also show for the one-component system the curve
for the onset of mechanical instability predicted by the C1
approximation [see Eq. (\ref{C1:mech})] and the one predicted
by the PY approximation [see Eq. (\ref{PY:mech})]. 

\item[Fig. \ref{fig:percolation-c1-poly}] Dependence of
the percolation threshold, as calculated from the C1 approximation
using Case V (see section \ref{sec:C1percolation}), from the
polydispersity.

\item[Fig. \ref{fig:binodal-c1}] Binodal curve and percolation
threshold [see Eq. (\ref{C1:percolation})], for a one-component
system, in the C1 approximation. For comparison 
we also show the percolation threshold of the Percus-Yevick
approximation \cite{Chiew83} (which exists for $\tau\ge 1/12$), the
one from the Monte Carlo simulation of Miller and Frenkel 
\cite{Miller04} (circles are the simulation results and the
fit, the dot-dashed line, is only valid for $\tau\ge 0.095$), the
binodal curve of the Percus-Yevick approximation (from the energy
route) \cite{Watts71}, and the binodal curve from the Monte Carlo
simulation of Miller and Frenkel \cite{Miller04} (points with
errorbars are the simulation results and the fit, the dot-dashed line,
is merely to guide the eye).

\item[Fig. \ref{fig:pd-m1}] Cloud and shadow curves for Case I in the 
mMSA at two  
values of polydispersity: $s=0.1$ and $s=0.3$. For the case $s=0.3$ we
also show three coexistence curves (continuous lines) obtained setting
$\rho^{(0)}=0.08$, $\rho^{(0)}=0.25$, and
$\rho^{(0)}=0.197\simeq\rho_{cr}$. For comparison the binodal of the
monodisperse $(s=0)$ system has also been included. 

\item[Fig. \ref{fig:dist-m1}] Evolution of the size distribution of
the coexisting phases 
$F^{(1)}(\sigma)$ and $F^{(2)}(\sigma)$, with temperature along the
critical binodal of Fig. \ref{fig:pd-m1} ($s=0.3$,
$\rho^{(0)}=0.197\simeq\rho_{cr}$). For comparison also the parent
Schulz distribution is shown (continuous line).

\item[Fig. \ref{fig:cs-s.3}] Cloud and shadow curves for
Case I, IV, and V in the mMSA at $s=0.3$. For comparison the binodal
of the monodisperse $(s=0)$ system has also been included (continuous
line). 
\end{itemize} 
\newpage
\centerline{\bf LIST OF TABLES}
\begin{itemize}

\item[Table \ref{tab:crit}] For the one-component system, we compare
the critical parameters obtained 
from the mMSA, C1, and PY \cite{Watts71} approximations with the ones
from the Monte Carlo simulation of Miller and Frenkel \cite{Miller04}.

\end{itemize} 
\newpage
\begin{table}[h!]
\begin{center}
\begin{tabular}{lcccccc}
\hline
&$~~~~~~~$& $\tau_c$ &$~~~~~~~$& $\eta_c$ &$~~~~~~~$& $(Z_{SHS})_c$\\
\hline
mMSA && 0.0943 && 0.13 && 0.36  \\
C1   && 0.1043 && 0.14 && 0.37  \\
PY   && 0.1185 && 0.32 && 0.32  \\
MC   && 0.1133 && 0.27 && -     \\
\hline
\end{tabular}
\caption[]{R. Fantoni, D. Gazzillo, and A. Giacometti
\label{tab:crit}}
\end{center}
\end{table}

\newpage

%

\end{document}